\documentclass{aa}  
\usepackage{natbib}
\usepackage{graphicx}
\usepackage{txfonts}

\newcommand{\kms}{{\hbox {\,km\,s$^{-1}$}}}
\newcommand{\mum}{{\hbox {\,$\mu$m}}}

\newcommand{\Lsun}{{\hbox {$L_\odot$}}}

\begin{document} 

  \title{NOEMA redshift measurements of bright {\it  Herschel} galaxies}
  \author{R. Neri\inst{1}          \and
          P. Cox\inst{2}           \and
          A. Omont\inst{2}         \and   
          A. Beelen \inst{3}       \and      
          S. Berta\inst{1}         \and  
          T. Bakx\inst{4,5,6}        \and 
          M. Lehnert\inst{2}       \and 
          A.~J. Baker\inst{7}        \and 
          V. Buat\inst{3}          \and 
          A. Cooray\inst{8}       \and 
          H. Dannerbauer\inst{9,10}   \and 
          L. Dunne\inst{4}         \and 
          S. Dye\inst{11}          \and
          S. Eales\inst{4}         \and
          R. Gavazzi\inst{2}       \and 
          A.~I. Harris\inst{12}    \and 
          C.~N. Herrera\inst{1}       \and 
          D. Hughes\inst{13}       \and 
          R. Ivison\inst{14,15}     \and 
          S. Jin\inst{9,10}            \and
          M. Krips\inst{1}         \and 
          G. Lagache\inst{3}       \and 
          L. Marchetti\inst{16,17,18}     \and  
          H. Messias\inst{19}      \and  
          M. Negrello\inst{4}      \and 
          I. Perez-Fournon\inst{9} \and 
          D.~A. Riechers\inst{20,21}     \and 
          S. Serjeant\inst{22}      \and 
          S. Urquhart\inst{22}   \and
          C. Vlahakis\inst{23}     \and 
          A. Wei{\ss}\inst{24}     \and 
          P. van der Werf\inst{25} \and 
          C. Yang\inst{26}         \and
          A.~J. Young\inst{7}
}         

  \institute{Institut de Radioastronomie Millim\'etrique (IRAM), 300 rue de la Piscine, 38400 Saint-Martin-d'H{\`e}res, France\\
              \email{neri@iram.fr}
          \and
             Sorbonne Universit{\'e}, UPMC Universit{\'e} Paris 6 and CNRS, UMR 7095, 
             Institut d'Astrophysique de Paris, 98bis boulevard Arago, 75014 Paris, France
          \and     
             Aix-Marseille Universit\'{e}, CNRS and CNES, Laboratoire d'Astrophysique de Marseille, 
             38, rue Frédéric Joliot-Curie 13388 Marseille, France
          \and
             School of Physics and Astronomy, Cardiff University, The Parade, Cardiff CF24 3AA, UK
          \and 
            Division of Particle and Astrophysical Science, Graduate School of Science, 
            Nagoya University, Aichi 464-8602, Japan
          \and  
            National Astronomical Observatory of Japan, 2-21-1, Osawa, Mitaka, Tokyo 181-8588, Japan
          \and 
             Department of Physics and Astronomy, Rutgers, The State University of New Jersey, 
             136 Frelinghuysen Road, Piscataway, NJ 08854-8019, USA
          \and
              University of California Irvine, Physics \& Astronomy, FRH 2174, Irvine CA 92697, USA
          \and 
             Instituto Astrof{\'i}sica de Canarias (IAC), E-38205 La Laguna, Tenerife, Spain  
          \and
             Universidad de La Laguna, Dpto. Astrofísica, E-38206 La Laguna, Tenerife, Spain
          \and 
             School of Physics and Astronomy, University of Nottingham, University Park, 
             Notthingham NG7 2RD, UK 
          \and
              Department of Astronomy, University of Maryland, College Park, MD 20742, USA
          \and
              Instituto Nacional de Astrofísica, \'Optica y Electr\'onica, Astrophysics Department, Apdo 51 y 216, Tonantzintla, Puebla 72000 Mexico
          \and 
             European Southern Observatory, Karl-Schwarzschild-Strasse 2, D-85748 Garching, Germany
          \and  
             Institut for Astronomy, University of Edinburgh, Blackford Hill, Edinburgh EH9 3HJ, UK 
          \and
             University of Cape Town, Department of Astronomy. Private Bag X3 Rondebosch, 7701 Cape Town, South Africa
           \and 
             Department of Physics and Astronomy, University of the Western Cape, Private Bag X17, Bellville 7535, Cape Town, South Africa 
           \and  
             Istituto Nazionale di Astrofisica, Istituto di Radioastronomia, via Gobetti 101, 40129 Bologna, Italy
           \and 
             Instituto de Astrofísica e Ciências do Espaço, Tapada da Ajuda, Edifício Leste, 1349-018 Lisboa, Portugal 
           \and
              Department of Astronomy, Cornell University, Space Sciences Building, Ithaca, New York (NY) 14853, USA
           \and
              Max-Planck-Institut f\"ur Astronomie, K\"onigstuhl 17, D-69117 Heidelberg, Germany
            \and
              Department of Physical Sciences, The Open University, Milton Keynes MK7 6AA, UK
           \and 
              National Radio Astronomy Observatory, 520 Edgemont Road, Charlottesville VA 22903, USA
           \and  
              Max-Planck-Institut f{\"u}r Radioastronomie, Auf dem H{\"u}gel 69, 53121 Bonn, Germany.
            \and 
              Leiden University, Leiden Observatory, PO Box 9513, 2300 RA Leiden, The Netherlands
            \and 
              European Southern Observatory, Alonso de C{\'o}rdova 3107, Casilla 19001, Vitacura, Santiago, Chile
         }

  \date{Received *** 2019/ accepted *** 2019}

 \abstract 
   {Using the IRAM NOrthern Extended Millimeter Array (NOEMA), we  conducted a program to measure redshifts 
   for 13 bright galaxies detected in the {\it Herschel} Astrophysical Large Area Survey 
   (H-ATLAS) with $\rm S_{500 \, \mu m} \geq 80 \, mJy$.  We report reliable spectroscopic redshifts for 
   12 individual sources, which are derived from scans of the 3 and 2 mm bands, covering up to 
   31 GHz in each band, and are based on the detection of at least two emission lines.
   The spectroscopic 
   redshifts are in the range  $2.08<z<4.05$ with a median value of $z=2.9\pm0.6$. The sources 
   are unresolved or barely resolved on scales of 10 kpc. In one field, two galaxies with different 
   redshifts were detected. In two cases  the sources are found to be binary galaxies with projected distances
   of $\rm \sim$140\,kpc. The linewidths of the sources are large, with a mean value for the 
   full width at half maximum of $\rm 700 \pm 300 \, km \, s^{-1}$ and a median of $\rm 800 \, km \, s^{-1}$. We analyze the nature 
   of the sources with currently available ancillary data to determine if they are lensed or hyper-luminous 
   ($L_{\rm FIR} > 10^{13} \, L_{\sun}$) galaxies. We also present a reanalysis of the spectral energy distributions 
   including the continuum flux densities measured at 3 and 2 mm to derive the overall properties of the sources. 
   Future prospects based on these efficient measurements of redshifts 
   of high-$z$ galaxies using NOEMA are outlined, including a comprehensive survey of all the brightest {\it Herschel} galaxies.}

\keywords{galaxies: high-redshift -- 
          galaxies: ISM  -- 
          gravitational lensing: strong -- 
          submillimeter: galaxies -- 
          radio lines: ISM}

\authorrunning{R. Neri et al.}

\titlerunning{NOEMA Redshift Measurements of Bright {\it Herschel} Galaxies}

\maketitle

\section{Introduction}
Some of the most vigorous  star formation activity occurred in submillimeter galaxies (SMGs) 
and other populations of dusty star-forming galaxies (DSFGs)
in the early universe \citep[see, e.g., reviews in][]{Blain2002, Casey2014}, whose rest-frame 8-1000\,$\mu$m luminosities ($L_{\rm IR}$) exceed a few $10^{12} \, \Lsun$. 
Their exact nature is still debated \citep[e.g.,][]{Narayanan2015},  
although many of them are probably mergers \citep[e.g.,][]{Tacconi2008}. Compared to local ultra-luminous infrared galaxies, 
SMGs are more luminous and several orders of magnitude more numerous. 
With a median redshift of $z \sim2.5$ \citep[e.g.,][]{Danielson2017}, SMGs
are most commonly found around the $z \sim2-3$ peak of the cosmic star formation rate density \citep{Madau2014}, and therefore play a critical 
role in the history of cosmic star formation as the locus of the physical processes driving the most extreme 
phases of galaxy formation and evolution. 

The SPIRE instrument \citep{Griffin2010} on the {\it Herschel Space Observatory} \citep{Pilbratt2010} 
has increased the number of known SMGs from hundreds to hundreds of 
thousands  through the {\it Herschel} Astrophysical Terahertz Large Area Survey 
\citep[H-ATLAS;][]{Eales2010}, covering an area of $\rm 616 \, deg^2$;
the {\it Herschel} Multi-tiered Extragalactic Survey \citep[HerMES;][]{Oliver2012}, 
covering an area of $\rm 430 \, deg^2$; and the {\it Herschel} Stripe 82 Survey \citep[HerS;][]{viero2014} covering an area of 81\,deg$^2$. 
As shown by \citet{Negrello2010}, the surface density of 
unlensed sources tends to zero around flux densities S$_{500\mu\rm m}$$\sim$100\,mJy, and 
most objects that are detectable above this threshold are gravitationally magnified by foreground galaxies. 
The South Pole Telescope (SPT) cosmological survey, covering an area of 2500~$\rm deg^2$, also revealed a 
significant population of strongly gravitationally lensed, high-redshift DSFGs  \citep[][]{Vieira2010, Spilker2016}. 
These and other large-area surveys, like the all-sky {\it Planck}-HFI, have therefore enabled the detection of numerous DSFGs that are among 
the brightest in the sky, 
including large fractions of the rare high-redshift strongly lensed systems 
\citep[][]{Negrello2010, Wardlow2013, Bussmann2013, Bussmann2015, Planck2015, Spilker2016, Nayyeri2016, Negrello2017, Bakx2018} 
and hyper-luminous infrared galaxies (HyLIRGs) with $L_{\rm FIR} > 10^{13} \, \Lsun$
\citep[see, e.g.,][]{Ivison2013, Fu2013, Oteo2016, Riechers2017}.

Exploiting this richness of data presents us with a tremendous task. In particular, precise spectroscopic measurements 
of the redshifts of individual sources are essential to derive their nature and  
physical properties, and to reveal their clustering characteristics, while photometric redshifts 
are only indicative of a redshift range \citep{Casey2012, Ivison2016}.
Conventional optical and near-infrared 
spectroscopy using large ground-based telescopes is possible for sources 
with precise positions available through their faint radio emission, but misses the dustiest 
bright objects and most of the highest redshift ($z>3$) sources, which lack radio counterparts \citep{Chapman2005}. 
Moreover, in the case of sources that are gravitationally amplified, the optical spectra 
detect, in most cases, the foreground lensing galaxies rather than the lensed objects.
(Sub)millimeter spectroscopy typically searches for CO emission lines, which are unhindered by dust 
extinction and can be related unambiguously to the (sub)millimeter source. 
It  therefore offers a far better alternative to the imprecise photometric method 
for deriving secure values for the redshifts.  

The spectroscopic method has only recently become competitive with the increased bandwidths of the receivers operating 
at millimeter and submillimeter facilities. Its power to reliably measure redshifts
was first   
demonstrated in the case of a few SMGs  detected by the Submillimetre Common-User Bolometer Array (SCUBA) in the continuum \citep{Smail1997, Hughes1998}. 
Their redshifts could only be determined more than a decade later, after various unsuccessful attempts, using the new 
broadband receivers that became available at the IRAM 30-meter telescope \cite[e.g., SMMJ14009+0252:][]{Weiss2009}  
at the Plateau de Bure interferometer \cite[e.g., HDF.850.1:][]{Walter2012} 
and at the Green Bank Telescope (GBT) for various SMGs 
\cite[e.g.,][]{Swinbank2010, Harris2010}. Subsequent broadband observations with the Zpectrometer on the 
GBT \citep{Harris2012}, with Z-Spec on the Caltech Submillimeter Observatory \citep{Lupu2012}, 
{ with the Combined Array for Research in Millimeter-wave Astronomy \cite[CARMA;][]{Riechers2011}, and recently with EMIR at the IRAM 30-meter telescope and VEGAS at the Green Bank Telescope (GBT; Bakx et al.\ in preparation)} enabled the measurement of 
tens of redshifts for very bright sources selected from the {\it Herschel} wide surveys. 

Using the Atacama Large Millimeter Array (ALMA), \citet{Weiss2013} presented the first redshift survey for 23 strongly lensed DSFGs selected 
from the SPT survey. This work was followed by further ALMA observations yielding reliable 
measurements for redshifts of an additional 15 DSFGs  from the 
SPT sample \citep{Strandet2016}. We note that the SPT-selected galaxies are at significantly higher 
redshifts (a median of $z \sim3.9$) than the {\it Herschel}-selected 
galaxies (mostly $2<z<3$ for the sources peaking in the 350~$\rm \mu m$ band), 
due to the difference in the frequency bands used in these surveys (see Sect.~\ref{section:spectroscopic-redshifts}). 
In parallel, a number of bright 
{\it Herschel} sources were observed by our team, with IRAM and other facilities including ALMA, yielding 
secure redshifts for about 50 sources \citep[see references in][]{Bakx2018, Nayyeri2016, Bussmann2013}.

The new NOEMA correlator, with its ability to 
process a total instantaneous bandwidth of 31 GHz in two frequency settings, alleviates one of the main 
problems  related to the measurement of redshifts of dust-obscured galaxies, namely the large overheads 
that are currently required in spectral-scan mode. We present here the results of a Pilot Program, whose  aim was   measuring redshifts for 13  bright SMGs (with $\rm S_{500\mu m} \ge 80 \, mJy$) selected 
from the H-ATLAS survey by performing 3 and 2 mm spectral scans. 
For 85\% of these H-ATLAS sources we obtain reliable redshifts based on the 
detection of CO emission lines at both 3 and 2 mm, demonstrating that NOEMA is able 
to efficiently measure redshifts of bright SMGs by scanning the 3 and 2 mm bands. 
This Pilot Program lays the ground work for a larger ongoing 
NOEMA program ({\it z-}GAL) that will derive spectroscopic redshifts for all the northern and equatorial 
bright $z \gtrsim2$ galaxies selected from the {\it Herschel} surveys (H-ATLAS, HerMES, and HerS) for 
which no reliable redshifts measurements are available.    
 
The structure of the paper is as follows. In Section~\ref{section:obs} we describe the sample 
selection, the observations, and the data reduction. In Section~\ref{section:results} we present the main results including 
the redshift determination, the spectral properties of the sources and their nature, and the properties of the 
continuum emission. In Section~\ref{section:discussion} we compare the spectroscopic and 
photometric redshifts, reassess the spectral energy 
distributions of the targets taking into account the continuum flux densities at 3 and 2 mm, derive
dust temperatures and infrared luminosities, discuss the widths of the CO emission lines, 
present the general properties of the sources (including CO luminosities and gas masses), and discuss 
the nature of each source, categorizing the lensed and hyper-luminous galaxies.  
Finally, in Section~\ref{section:conclusions} we summarize the main conclusions and outline future prospects.   
 
Throughout this paper we adopt a spatially flat $\Lambda$CDM cosmology 
with $H_{0}=67.4\,{\rm km\,s^{-1}\,Mpc^{-1}}$ and $\Omega_\mathrm{M}=0.315$ \citep{Planck2018}.
 

\begin{figure*}
   \centering
\includegraphics[width=0.9\textwidth]{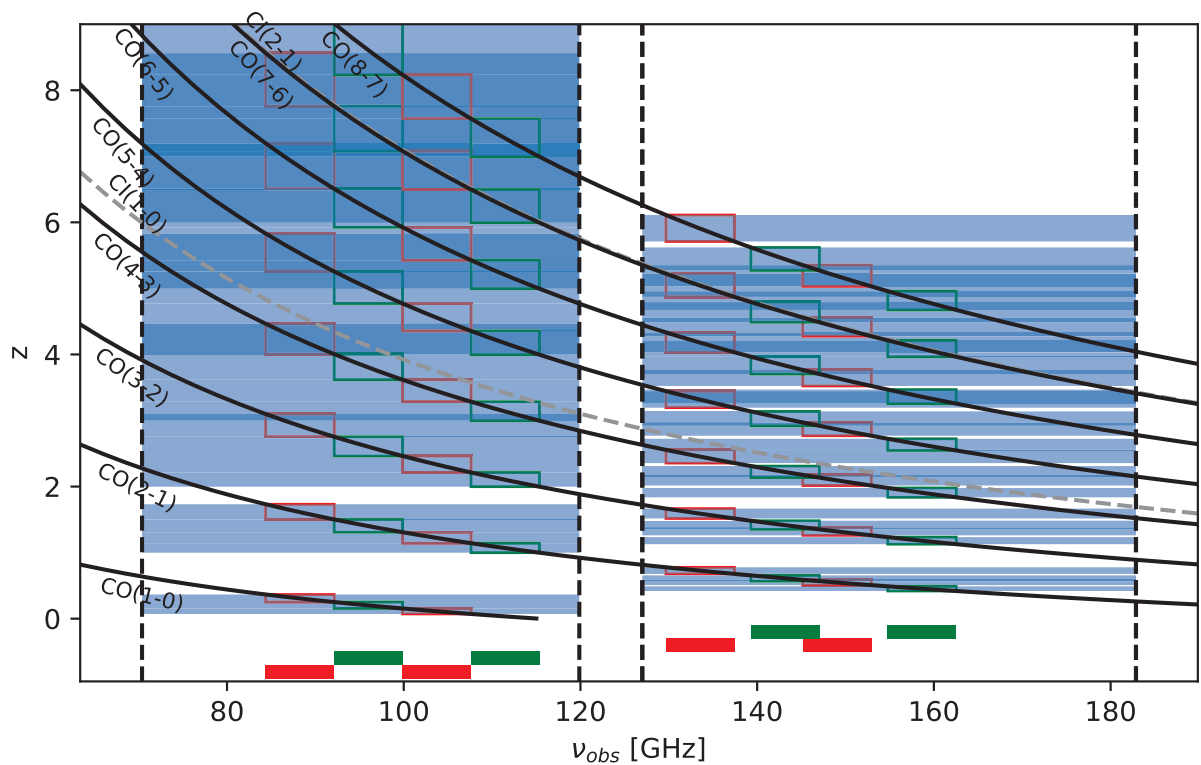}
   \caption{Spectral coverage of the $^{12}$CO (solid black) and $\rm [C{\small I}]$ (dashed gray) emission lines as a function of redshift in the
   3 and 2 mm atmospheric windows in the frequency ranges 84-115~GHz and 130-163~GHz. The bottom colored boxes show 
   the LSB and USB frequency settings (red and green, respectively) (see Table\,\ref{table:Observing_Log}).  The 2 mm frequency windows were selected to optimally cover the range of spectroscopic redshifts predicted by the 3 mm observations.  
   The dark blue zones identify the redshift ranges where at least two emission lines 
   are detected at 3 or 2 mm with the current settings, while the light blue zones indicate the redshift 
   ranges where only one line is present. This wide frequency range enables the detection of at least one emission line in each band, 
   except for a few small redshifts gaps (see Sect.~\ref{section:obs} for details). }
   \label{figure:spectral-coverage}
\end{figure*}

\begin{figure}
   \centering
\includegraphics[angle=270,width=0.45\textwidth]{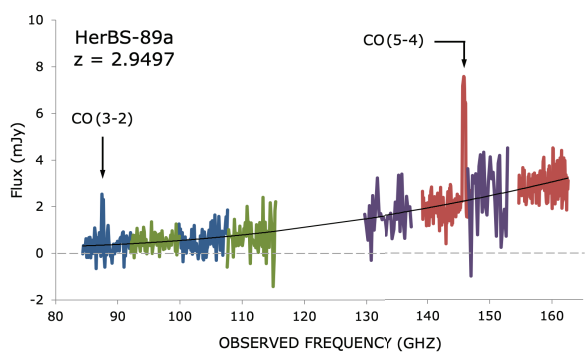}
   \caption{Spectral setup and frequency coverage of the frequency settings in the NOEMA 3 and 2 mm bands for HerBS-89a (see Table~\ref{table:Observing_Log}). The four settings are shown in different colors. The $^{12}$CO emission lines detected in HerBS-89a are identified and the solid line is a fit to the underlying dust continuum (see Sect.~\ref{section:individual-sources} for details on the source).}
   \label{figure:set-up}
\end{figure}

\section{Sample selection and observations}
\label{section:obs}

\subsection{Sample selection}
The 13 sources of the Pilot Program were selected from the {\it Herschel} Bright Sources (HerBS) 
sample, which contains the 209 galaxies detected in the H-ATLAS survey with $\rm S_{500 \mu m}>80$\,mJy 
and photometric redshifts $z_{\rm phot} > 2$ \cite[]{Bakx2018}. Most of these galaxies 
have been observed at 850\,$\mu$m using SCUBA-2, and only 22 sources have spectroscopic 
redshifts \citep[see references in][]{Bakx2018}. We note  that the SCUBA-2 flux densities originally reported in 
\cite{Bakx2018} have  recently been revised using the method described in \cite{Ivison2016} and \cite{duivenvoorden2018}, together with the estimated photometric redshifts, as explained in an Erratum to that paper (Bakx et al., in preparation). The SCUBA-2 flux densities and the photometric redshifts $z_{\rm phot}$ listed in Table~\ref{table:sources} are the revised values. 

The selected galaxies are located in the largest wide field observed by {\it Herschel} in the northern sky, 
in the vicinity of the North Galactic Pole (NGP). Measuring $\rm 15 \times 10\,deg^2$ and centered on 
[R.A.=13$\rm ^h$, Dec.=29\,deg], the declination of the NGP field is optimal for NOEMA observations, 
and its size allowed us to group sources during the observations, minimizing per-source overheads 
(see §2.2). The NGP field contains 49 high-$z$ sources 
with $\rm S_{500 \, \mu m}>80$\,mJy; from this list, we extracted 
13 sources for which no spectroscopic redshift measurements were available. Six have 
$\rm S_{500 \, \mu m} > 100 \, mJy$, completing the redshift determination for sources with 
$\rm S_{\rm 500 \, \mu m} > 100 \, mJy$ (in the NGP field) at $z_{\rm phot} > 2$. 
The selected galaxies are therefore in the range 2 $\lesssim z_{\rm phot}\lesssim 3$, with flux densities 
in the range $\rm 80\, mJy  \lesssim S_{500 \, \mu m} \lesssim 130 \, mJy$ (Table~\ref{table:sources}) 
and apparent far-infrared luminosities in excess 
of $10^{13} L_{\sun}$ (see Sect.~\ref{section:results} 
and Table~\ref{tab:source-properties}).

\begin{table*}[!htbp]
\tiny
\caption{The sample}
\begin{tabular}{lcccccc}
\hline\hline
\multicolumn{2}{c}{Source Name} & $z_{\rm phot}$ & \multicolumn{3}{c}{{\it Herschel} Flux Density} & SCUBA-2 Flux Density \\
HerBS & H-ATLAS &   & $\rm S_{250 \, \mu m}$ & $\rm S_{350 \, \mu m}$ & $\rm S_{500 \, \mu m}$ & $\rm S_{850 \, \mu m}$  \\
& & & \multicolumn{3}{c}{(mJy)} & (mJy) \\
\hline
HerBS-34  & J133413.8+260458 & 2.35 & 136.1$\pm$5.4 & 161.0$\pm$5.5 & 126.5$\pm$6.8 & 34.0$\pm$5.7 \\ 
HerBS-43  & J132419.0+320752 & 3.08 & 84.4$\pm$4.9 & 116.0$\pm$5.2  & 115.4$\pm$6.3 & 37.0$\pm$5.1 \\ 
HerBS-44  & J133255.8+342208 & 1.83 & 164.3$\pm$5.8 & 186.8$\pm$5.8 & 114.9$\pm$7.2 & 25.3$\pm$4.5 \\ 
HerBS-54  & J131540.6+262322 & 2.95 & 94.0$\pm$5.7 & 116.1$\pm$6.1  & 108.6$\pm$7.1 & 44.7$\pm$4.6 \\ 
HerBS-58  & J130333.1+244643 & 2.53 & 99.0$\pm$5.5 & 111.5$\pm$5.9  & 104.5$\pm$7.1 & 30.5$\pm$5.0 \\ 
HerBS-70  & J130140.2+292918 & 2.08 & 119.6$\pm$5.8 & 136.8$\pm$5.8 & 100.0$\pm$7.1 & 21.9$\pm$5.5  \\ 
HerBS-79  & J131434.1+335219 & 2.36 & 103.4$\pm$5.6 & 115.3$\pm$6.0 &  97.9$\pm$7.3 & 28.5$\pm$5.0  \\ 
HerBS-89  & J131611.5+281219 & 3.53 & 71.8$\pm$5.7 & 103.4$\pm$5.7  &  95.7$\pm$7.0 & 52.8$\pm$4.3  \\ 
HerBS-95  & J134342.5+263919 & 3.20 & 61.9$\pm$5.7 & 101.3$\pm$5.7  &  94.7$\pm$7.6 & 27.4$\pm$6.2  \\ 
HerBS-113 & J131211.5+323837 & 2.77 & 80.7$\pm$5.9 & 103.4$\pm$6.0  &  92.0$\pm$7.0 & 32.0$\pm$5.2  \\ 
HerBS-154 & J132258.2+325050 & 2.63 & 79.1$\pm$5.6 & 87.9$\pm$5.9   &  85.6$\pm$7.2 & 28.8$\pm$4.2  \\ 
HerBS-173 & J131804.7+325016 & 2.38 & 73.3$\pm$5.6 & 92.7$\pm$6.0   &  83.3$\pm$7.2 & 18.8$\pm$4.3  \\ 
HerBS-204 & J132909.5+300957 & 3.61 & 57.9$\pm$5.5 & 95.3$\pm$6.1   &  80.1$\pm$7.1 & 40.0$\pm$6.6  \\ 
\hline
\end{tabular}
 \tablefoot{The source names and {\it Herschel} flux densities are from \citet{Bakx2018}. The SCUBA-2 flux densities and  
 photometric redshifts ($z_{\rm phot}$) have been updated from that paper 
 based on a revision of the SCUBA-2 flux densities (see text); further details are provided in Bakx et al. (in preparation).}
   \label{table:sources}
\end{table*}
\normalsize 

\begin{table*}[!htbp]
\tiny       
\caption{Observation log}             
\begin{tabular}{lcccccc}  
\hline\hline       
Freq. Setting & 1 (Apr 18) & 1 (Aug 6/Aug 7) & 2 (Apr 21/Apr 24) &  3 (May 24) & 4 (Sep 8/Oct 19-23) & Total        \\
LSB-range & 84.385$-$92.129\,GHz & 84.385$-$92.129\,GHz &  92.129$-$99.873\,GHz  & 129.680$-$137.424\,GHz & 139.293$-$147.037\,GHz & $ t_{\rm obs}$ (min) \\
USB-range & 99.873$-$107.617\,GHz &99.873$-$107.617\,GHz & 107.617$-$115.361\,GHz & 145.168$-$152.912\,GHz & 154.781$-$162.525\,GHz &   \\
Configuration & 9C & 9D & 9C & 8D & 9D (Sep 8), 9C& \\
\hline
\hline
 & $t_{\rm obs}$, B$_{\rm L}$, B$_{\rm U}$ (") & $t_{\rm obs}$, B$_{\rm L}$, B$_{\rm U}$ (") & $t_{\rm obs}$, B$_{\rm L}$, B$_{\rm U}$ (") & $t_{\rm obs}$, B$_{\rm L}$, B$_{\rm U}$ (") & $t_{\rm obs}$, B$_{\rm L}$, B$_{\rm U}$ (") & \\
HerBS-34         & 11.7, 2.2$\times$2.1, 1.8$\times$1.6  & & 6.0, 2.4$\times$2.1, 2.1$\times$1.8 &                      & 89.1, 1.4$\times$1.3, 1.3$\times$1.2 & 106.8 \\
HerBS-43         & 11.7, 2.1$\times$1.9, 1.8$\times$1.6  & & 11.7, 2.1$\times$1.8, 1.9$\times$1.5  &  4.4, 3.7$\times$3.0, 3.3$\times$2.8 &  & 27.8 \\
HerBS-44 & 11.7, 2.1$\times$1.9, 1.8$\times$1.5 & & &  4.3, 3.6$\times$2.8, 3.2$\times$2.5 &    & 16.0                  \\
HerBS-54         & 11.7, 2.3$\times$2.0, 1.8$\times$1.6  & &                      &  5.1, 4.0$\times$3.2, 4.0$\times$3.2& & 16.7 \\
HerBS-58         & 11.7, 2.2$\times$1.9, 1.7$\times$1.7  & & 17.7, 2.1$\times$2.0, 2.0$\times$1.5  &  5.1, 4.2$\times$3.4, 3.8$\times$3.0 &  58.5, 1.5$\times$1.3, 1.3$\times$1.2 & 92.9 \\
HerBS-70         & 11.7, 2.1$\times$1.9, 1.8$\times$1.6  & 49.7, 6.9$\times$4.9, 5.8$\times$4.0 &                     &  5.1, 3.8$\times$3.3, 3.4$\times$3.0& 89.0, 1.5$\times$1.3, 1.3$\times$1.2  & 105.7 \\
HerBS-79 & 11.7, 2.1$\times$1.9, 1.8$\times$1.6 & & 18.4, 2.1$\times$1.8, 1.8$\times$1.5     
& 4.3, 3.6$\times$3.0, 3.2$\times$2.7 & &  34.4      \\
HerBS-89         & 11.7, 2.1$\times$1.9, 1.8$\times$1.6  & & 11.7, 2.4$\times$1.8, 2.0$\times$1.5 &  5.1, 3.9$\times$3.4, 3.5$\times$3.0&  59.3, 1.5$\times$1.3, 1.3$\times$1.2 & 87.7 \\
HerBS-95         & 11.7, 2.2$\times$1.9, 1.8$\times$1.5  & & 17.7, 2.1$\times$1.8, 1.9$\times$1.5 &  5.1, 3.9$\times$3.2, 3.5$\times$2.9& 88.5, 1.4$\times$1.3, 1.3$\times$1.2 & 122.9 \\
HerBS-113        & 11.7, 2.1$\times$1.9, 1.8$\times$1.6  & & 17.7, 2.2$\times$1.8, 1.9$\times$1.5 &  5.1, 3.6$\times$3.5, 3.3$\times$3.2&  & 34.4 \\
HerBS-154        & 11.7, 2.1$\times$1.9, 1.7$\times$1.7  &  &                     &  5.1, 3.6$\times$3.5, 3.2$\times$3.1& 89.0, 1.5$\times$1.3, 1.3$\times$1.2 & 105.7 \\
HerBS-173        & 11.7, 2.1$\times$1.9, 1.8$\times$1.6  & & 17.7, 2.2$\times$1.8, 1.8$\times$1.5  &                      &   & 29.3 \\
HerBS-204        & 11.7, 2.1$\times$1.9, 1.7$\times$1.6  & 45.6, 5.2$\times$4.0, 4.2$\times$3.2  & 17.7, 2.2$\times$1.8, 1.9$\times$1.5 &                      &    & 75.0 \\
\hline
\end{tabular}
\tablefoot{$t_{\rm obs}$ is the effective on-source integration time for the nine-element NOEMA array; 
a multiplicative factor of 1.6 should be used to estimate the total telescope time (i.e., including overheads). B$_{\rm L}$ and B$_{\rm U}$ are the synthesized beams at the center frequencies 
of the LSB and USB sidebands using natural weighting. HerBS-70 was observed on Aug 6 and 7, 2019 with 
the phase reference position placed midway (13:01:39.83 $+$29:29:20.8, J2000) between HerBS-70E and HerBS-70W. 
Observations of HerBS-204, made on Aug 6 and 7, 2019, were merged with data obtained on Apr 18, 2018.} 
\label{table:Observing_Log}
\end{table*}

\subsection{Observations}
We used NOEMA to observe the 13 selected bright SMGs (see Table~\ref{table:sources}) in the NGP field and derive their  
redshifts by scanning the 3 and 2 mm bands to search for at least two emission lines. 
The observations were carried out under projects W17DM and S18CR (PI: A.Omont) in the 
3 mm band with nine antennas, between April 18 and 24, 2018, and on August 6 and 7, 2019, 
and in the 2 mm band with eight and nine antennas, between May 24 and October 23, 2018. 
Observing conditions were on average excellent with an atmospheric phase stability of typically 
10-40\,deg RMS and 2-5\,mm of precipitable water vapor. The correlator was operated in the 
low-resolution mode to provide spectral channels with a nominal resolution of 2\,MHz. The observation log is presented in Table~\ref{table:Observing_Log}.

The NOEMA antennas are equipped with 2SB receivers that cover a spectral window
of 7.744\,GHz in each sideband and polarization. Since the two sidebands are separated by 7.744~GHz, two frequency settings are necessary to span a contiguous spectral window of 31\,GHz. At 3 mm, we adjusted the 
spectral survey to cover a frequency range from 84.385 to 115.361~GHz
(Table~\ref{table:sources}). At 2 mm, we then selected two frequency windows that  covered 
as well as possible the candidate redshifts allowed by the emission lines detected at 3 mm (see Table~\ref{table:Observing_Log}).  
The wide spectral coverage of the NOEMA correlator ensures that a scan of both the 3 and 2 mm spectral windows
can detect for every $z \lesssim4$ source at least one CO emission line, 
between $\rm ^{12}CO$\,(3-2) and $\rm ^{12}CO$\,(6-5), in each band, with the exception of a 
few redshift gaps. The gaps most relevant to the present observations are at 3 mm for $1.733<z<1.997$, 
and at 2 mm for $1.668<z<1.835$. (Fig.~\ref{figure:spectral-coverage}). The redshift range $1.733<z<1.835$ was not covered by any of the 3 and 2 mm settings. 
The spectral coverage of these observations also includes the 
$\rm [C{\small I}]\,(^3P_1$-$\rm ^3P_0)$ fine-structure line (492\,GHz rest-frame)
and, for sources at $z>3.65$, the water para-H$_2$O\,(2$_{11}$-$2_{02})$ transition (752\,GHz rest-frame), 
both of which were detected in the sources selected for this study (see Sect.~\ref{section:individual-sources} 
and Table~\ref{table:emission-lines}). Based on the redshift range of the sources, other lines of
abundant molecules are expected within the frequency range that was surveyed, such as 
HCO$^+$, HCN, or CN \citep[see, e.g.,][]{Spilker2014}; however, no further emission line, in addition to the
atomic carbon and water lines, was detected at the current sensitivity of the observations.
Exploring both the 3 and 2 mm spectral bands is therefore a prerequisite for 
detecting at least two CO emission lines in $2<z<4$ {\it Herschel}-selected bright galaxies, such as those 
selected for the Pilot Program, and deriving reliable spectroscopic redshifts. 
 
All observations were carried out in track-sharing mode by cyclically switching between galaxies within 
a track, as was possible due to the proximity of the sources. 
Two different configurations (C and D) of the array were used, yielding angular resolutions between
$1\farcs2$ and $3\farcs5$ at 2 mm, and $1\farcs7$ and $\sim$6$''$ at 3 mm. Observations were started by observing all 13 sources in one 
track in the lower 3 mm frequency setting. Galaxies that did not show a robust line detection were then 
observed again in the upper 3 mm frequency setting. For every line detection, the most probable redshift 
was estimated taking into account the photometric redshift. The galaxies were subsequently
observed in one of the two 2 mm frequency settings, and when the line was not detected,
observed again in the second setting. 
One example of the frequency settings in the 3 and 2 mm bands is shown in Fig.~\ref{figure:set-up} for the source HerBS-89a. 

For all the sources, the phase and amplitude calibrator was 1328+307 and the flux calibrators
MWC349 and LkH$\alpha$101. The data were calibrated, averaged in polarization, mapped, and analyzed in the GILDAS software package. 
The absolute flux calibration was estimated to be accurate to within 10\%. 
Source positions are provided with an accuracy of 0$\farcs2$ (Table~\ref{tab:flux-densities}).

\begin{figure*}
   \centering
\includegraphics[width=0.7\textwidth]{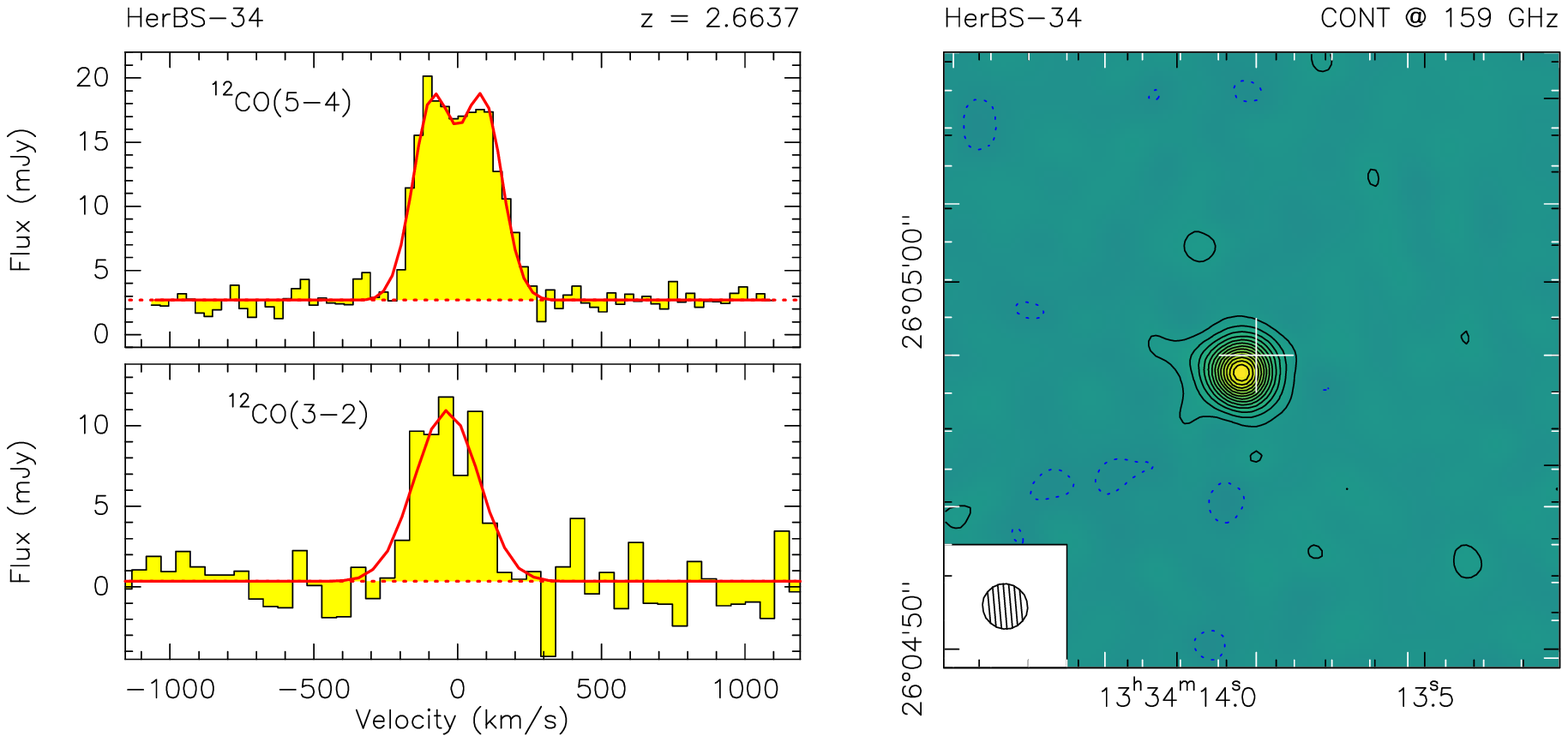} 
\includegraphics[width=0.7\textwidth]{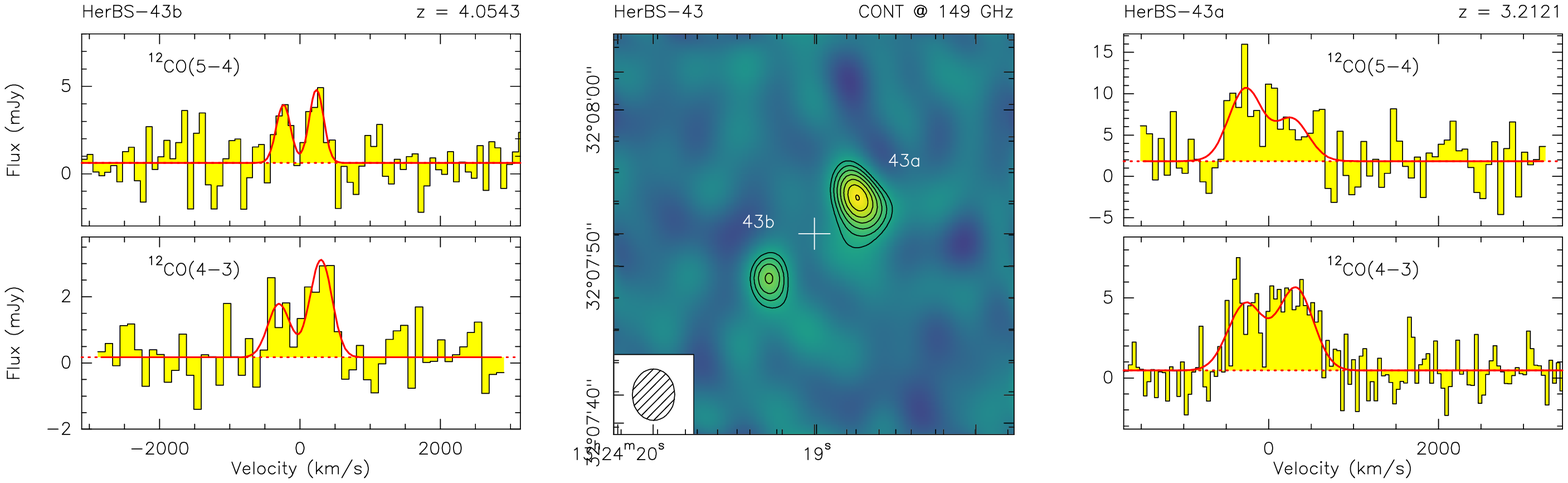} 
\includegraphics[width=0.7\textwidth]{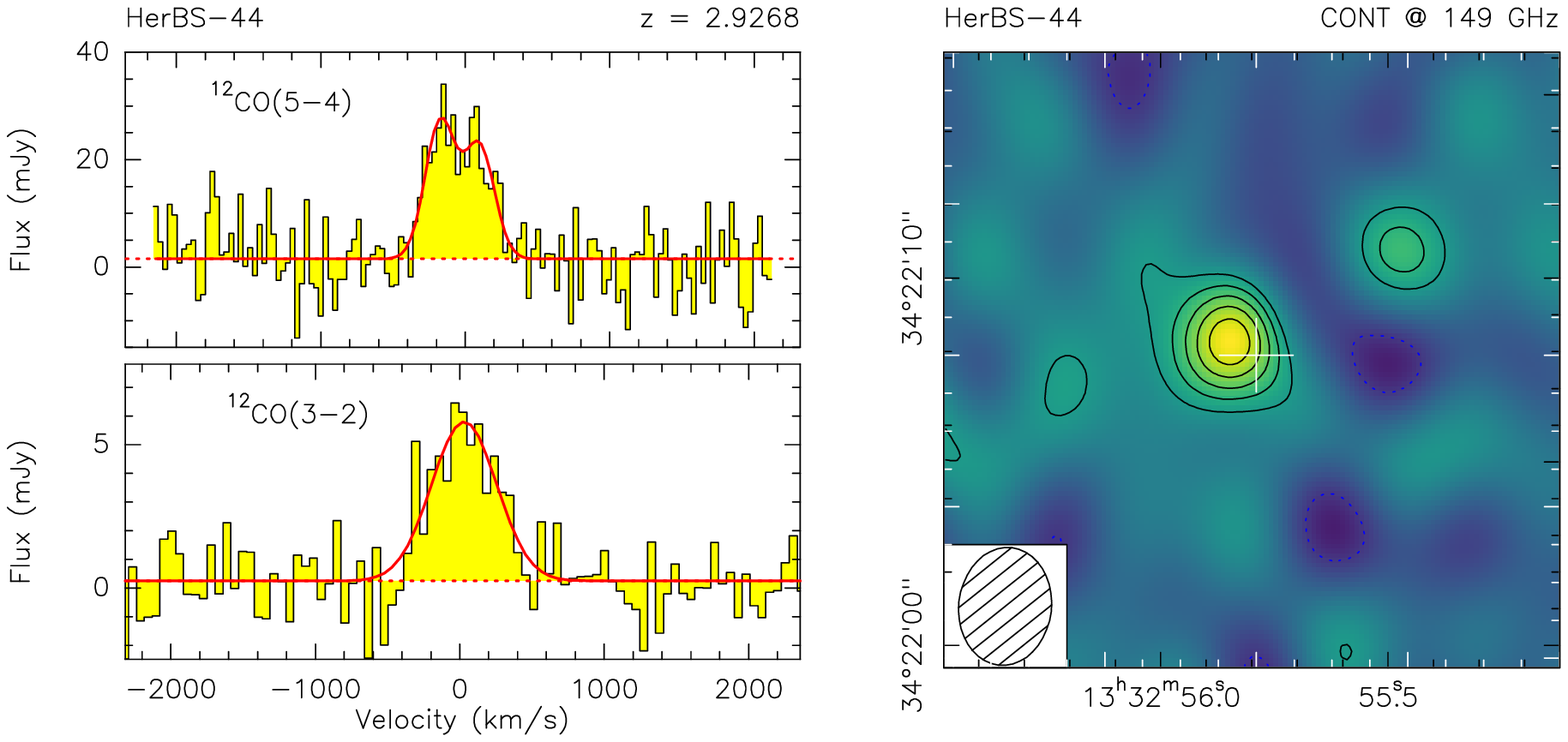} 
   \caption{Continuum images at 2 mm and spectra from the 2 mm (top) and 3 mm (bottom) bands for the {\it Herschel} bright 
            galaxies HerBS-34, HerBS-43, and HerBS-44.
            The source name, the continuum frequency, and the derived spectroscopic redshift are 
            indicated along the top of each panel. The emission lines are identified in the upper left 
            corner of each spectrum. The spectra are displayed with the continuum and each emission line is centered at the zero
            velocity corresponding to its rest frequency. 
            Fits to the continuum and the emission line profiles are shown as dotted and solid red lines, respectively.
            Continuum contours are plotted starting at $3\sigma$ in steps of $5\sigma$ and $1\sigma$ for HerBS-34 [42] and 
            HerBS-43 [284], respectively, and $2\sigma$ in steps of $1\sigma$ for HerBS-44 [283], where the numbers in brackets 
            are the $1\sigma$ noise levels for each source in $\rm \mu Jy \, beam^{-1}$. 
            In the case of HerBS-43, the panels showing the emission lines 
            on the left correspond to the source HerBS-43b, whereas the panels to the right 
            show the spectra of HerBS-43a. The synthesized beam is shown in the lower left corner of each continuum image.}
   \label{figure:spectra1}%
    \end{figure*}
    
    \begin{figure*}
   \centering
\includegraphics[width=0.7\textwidth]{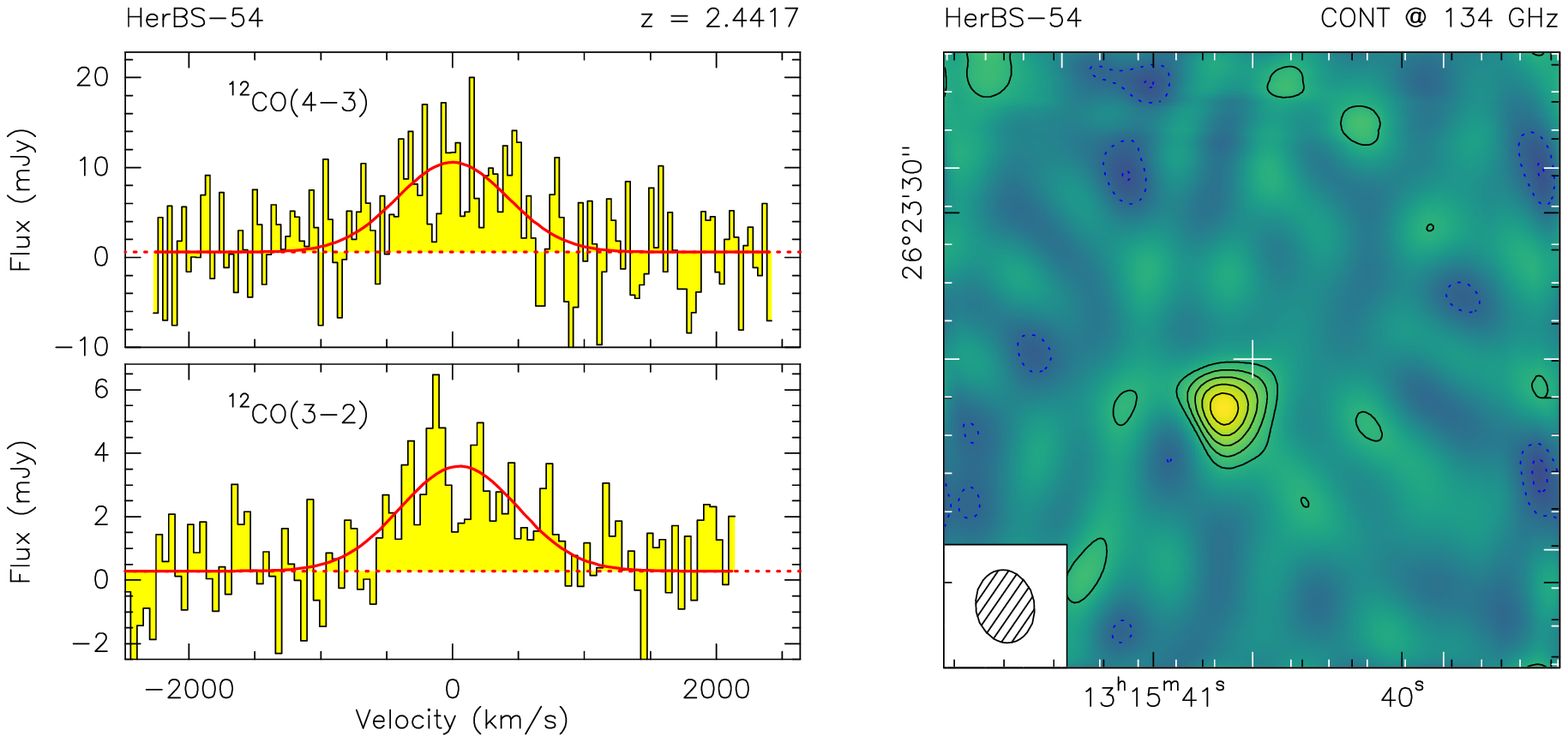}
\includegraphics[width=0.7\textwidth]{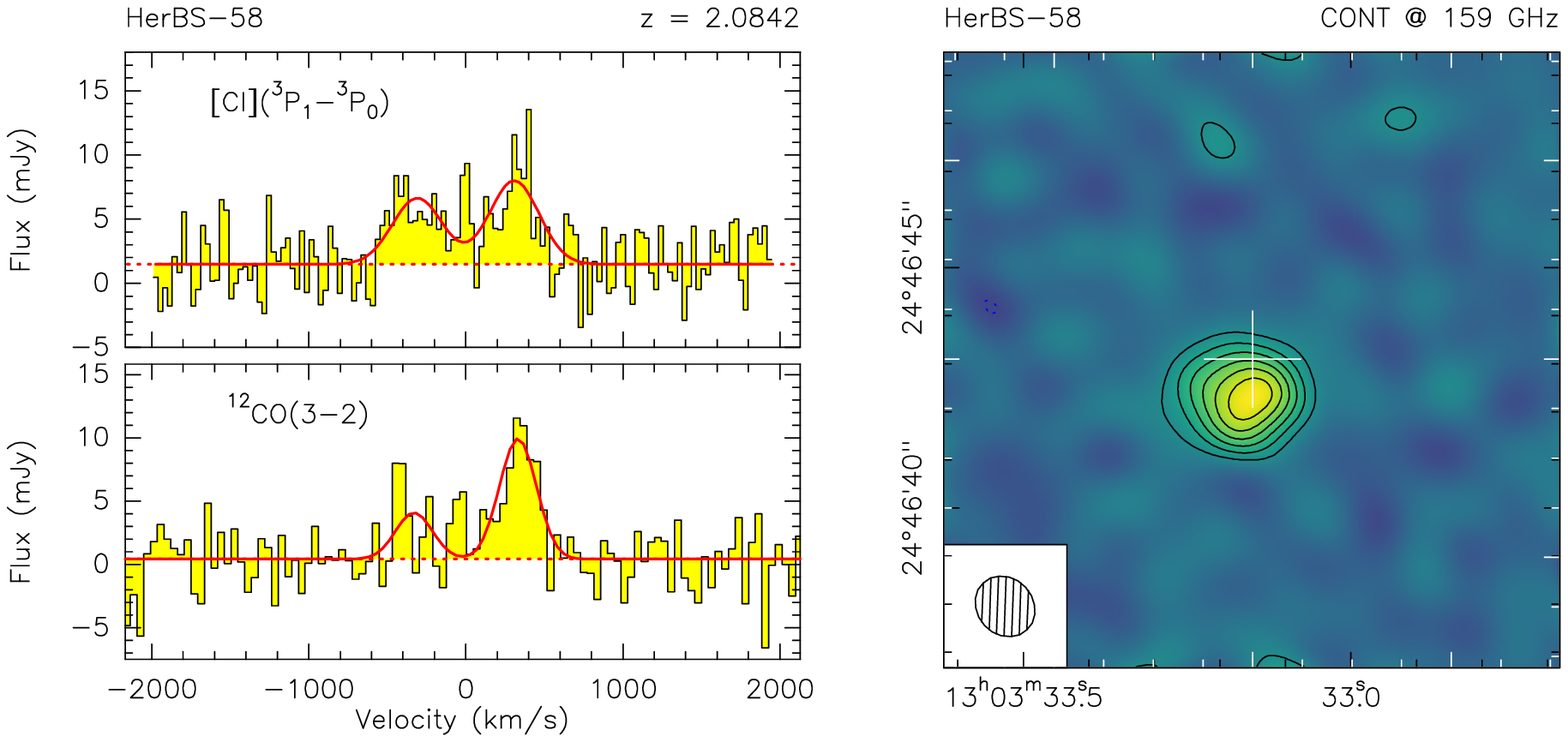}
\includegraphics[width=0.7\textwidth]{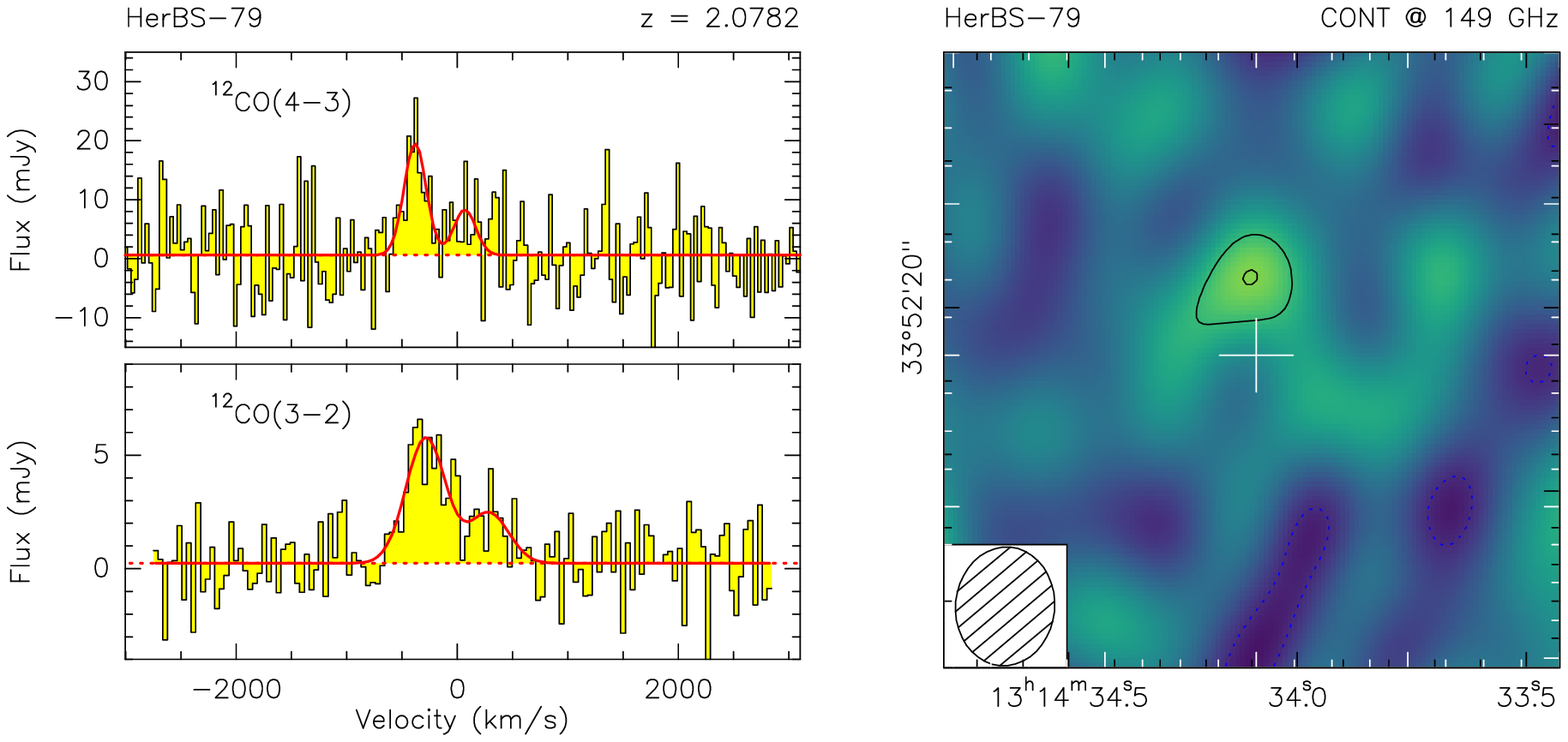}
   \caption{Continuum images at 2 mm (right) and spectra from the 2 mm (top) and 3 mm (bottom) bands of the {\it Herschel} bright 
            galaxies HerBS-54, HerBS-58, and HerBS-79.
            Continuum contours are plotted starting at $2\sigma$ in steps of $1\sigma$ for HerBS-54 [232], 
            $3\sigma$ in steps of $2\sigma$ for HerBS-58 [52], and $2\sigma$ in steps 
            of $1\sigma$ for HerBS-79 [297], where the numbers in brackets are the local noise levels $\sigma$ for each source 
            in $\rm \mu Jy \, beam^{-1}$.  See caption of Fig.\ref{figure:spectra1} for further details. }
   \label{figure:spectra2}    
    \end{figure*}
    
     \begin{figure*}
   \centering
\includegraphics[width=0.7\textwidth]{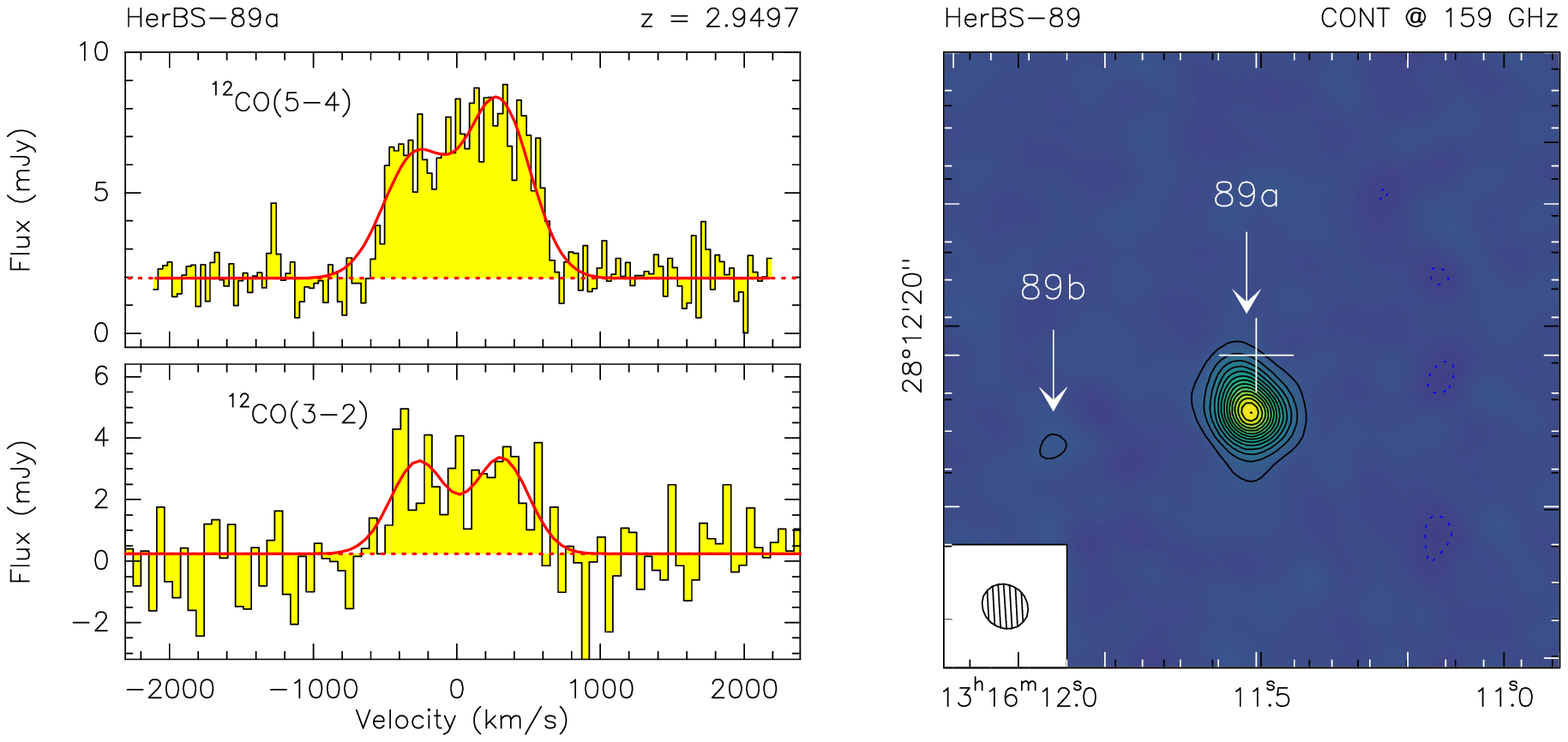}
\includegraphics[width=0.7\textwidth]{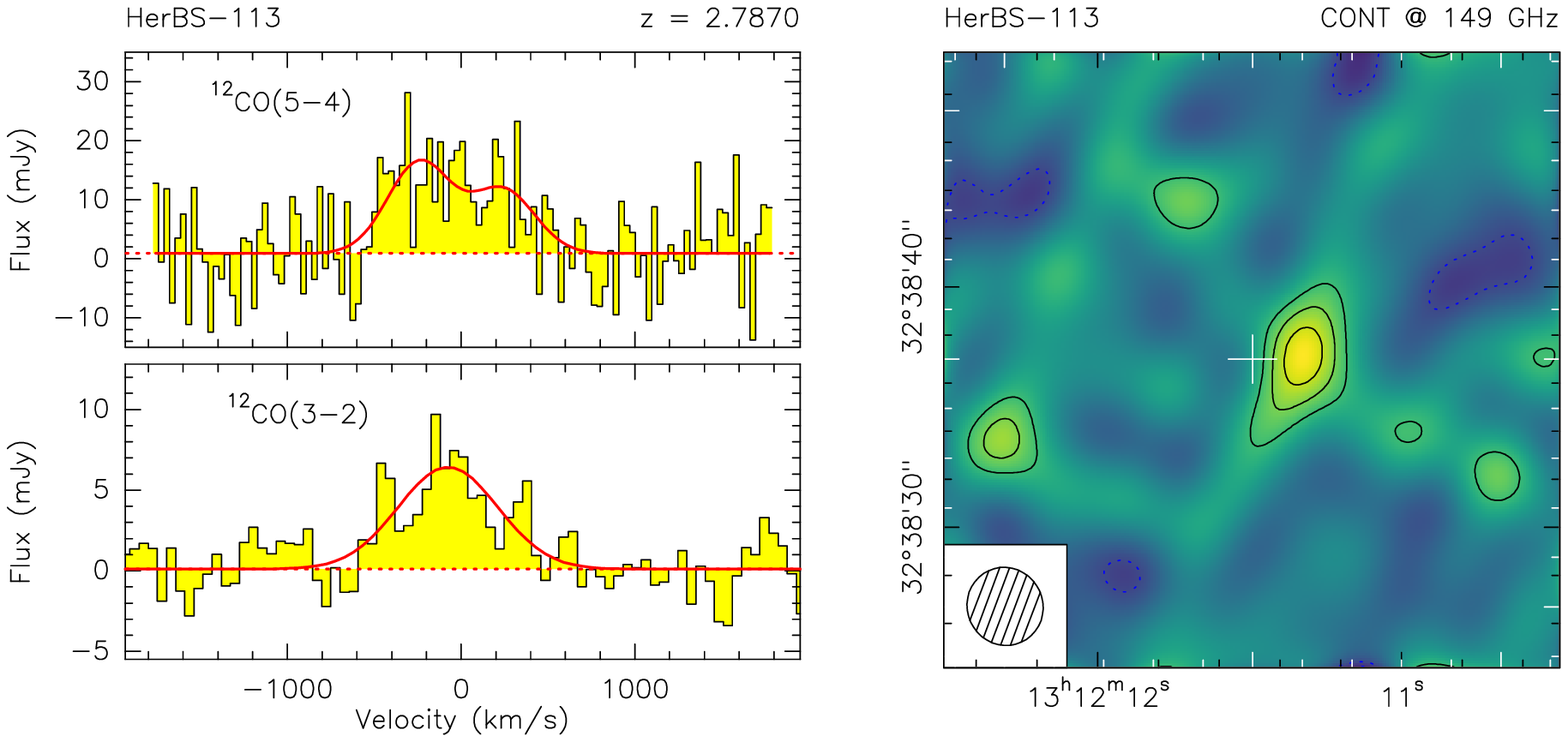}
\includegraphics[width=0.7\textwidth]{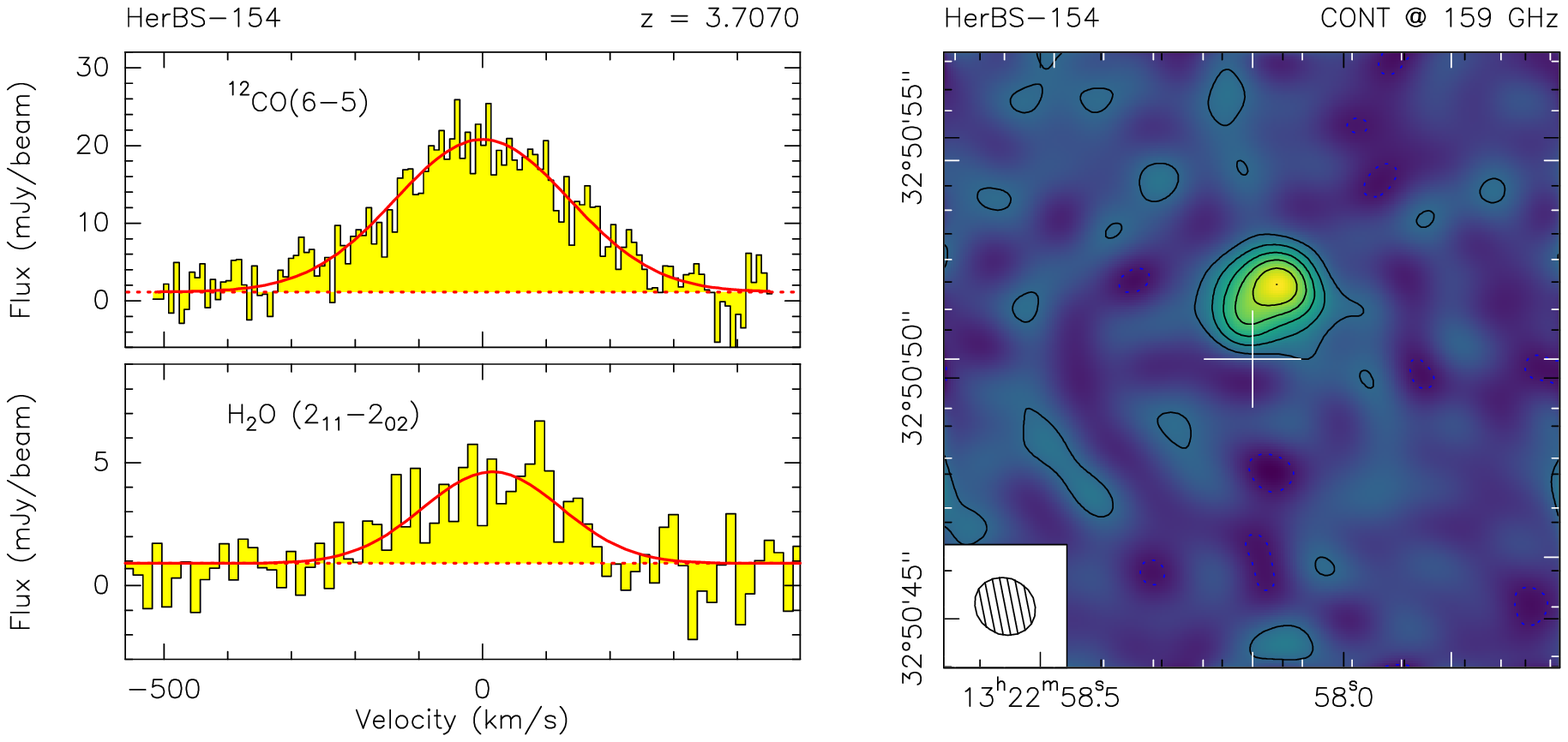}
   \caption{Continuum images at 2 mm (right) and spectra from the 2 mm (top) and 3 mm (bottom) bands of the Herschel bright 
            galaxies HerBS-89, HerBS-113, and HerBS-154. 
            In the case of HerBS-154, one of the emission lines shown is $\rm H_2O\,(2_{11}$-$2_{02})$. Continuum contours are 
            plotted starting at $3\sigma$ in steps of $5\sigma$ for HerBS-89 [49], 
            and $2\sigma$ in steps of $1\sigma$ and $2\sigma$ for HerBS-113 [308] and 
            HerBS-154 [82], respectively, where the numbers in brackets are the  local noise levels $\sigma$ for each source 
            in $\rm \mu Jy \, beam^{-1}$. See caption of  Fig.\ref{figure:spectra1} for further details.}
   \label{figure:spectra3}%
    \end{figure*}
      
    \begin{figure*}
   \centering
\includegraphics[width=0.8\textwidth]{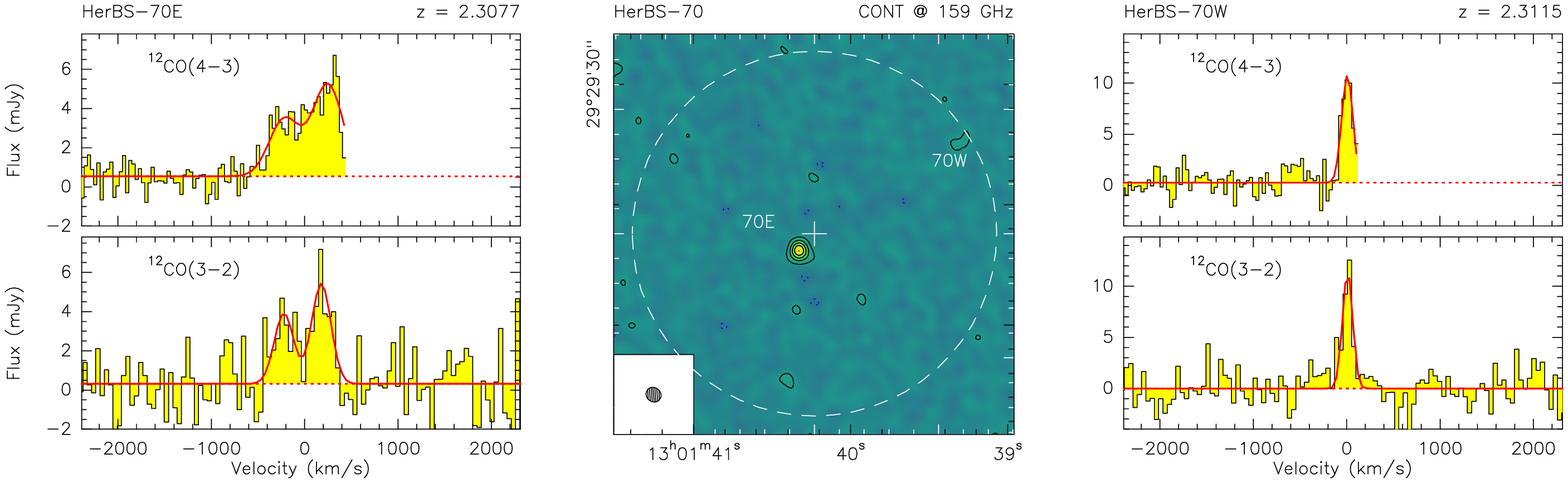}
\includegraphics[width=0.8\textwidth]{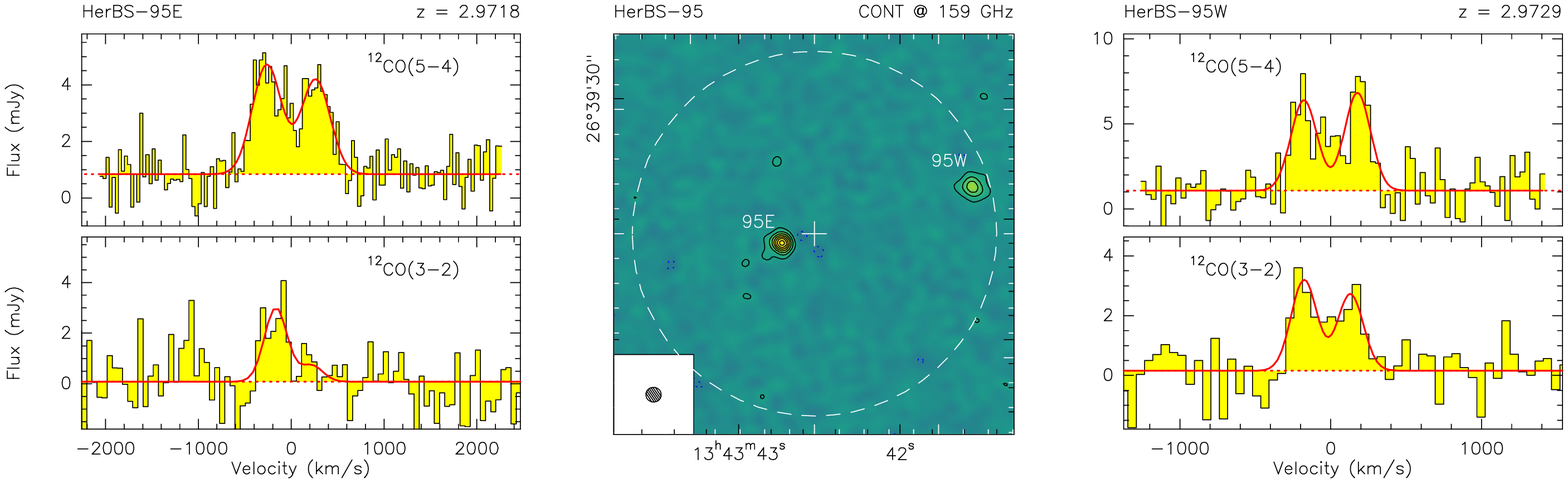}
   \caption{Continuum images at 2 mm (center) and spectra (left and right panels) from the 3 and 2 mm bands
            for the {\it Herschel} bright galaxies with a second source in the field at the same redshift: 
            HerBS-70 at $z=2.31$ (top panel) and HerBS-95 at $z=2.97$ (bottom panel) (see text for details). 
            The primary beam at 50\% is shown with a dashed circle. The panels showing the emission 
            lines on the left correspond to the (a) sources near the phase centers, whereas the panels to the right 
            show the spectra of the (b) companion sources to the west. Continuum contours are plotted 
            starting at $3\sigma$ in steps of $5\sigma$ for both  HerBS-70 [40] and HerBS-95 [43], 
            where the numbers in brackets are the local noise levels $\sigma$ for each source in $\rm \mu Jy \, beam^{-1}$. 
            The spectra are primary beam corrected (see caption of Fig.\ref{figure:spectra1} 
            for further details).} 
   \label{figure:spectra4}%
    \end{figure*}
 
 \begin{figure}
     \centering
     \includegraphics[width=0.35\textwidth]{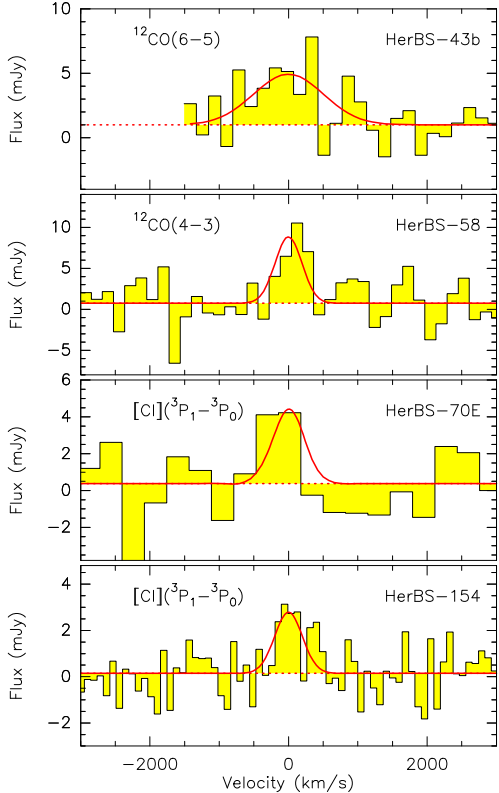}
     \caption{Other emission lines detected in the sources displayed in Figs.~\ref{figure:spectra1} 
              to \ref{figure:spectra4}. From top to bottom: HerBS-43b in $^{12}$CO\,(6-5); HerBS-58 in $^{12}$CO\,(4-3); 
              HerBS-70E and HerBS-154 in $\rm [C{\small I}]\,(^3$P$_1$-$^3$P$_0)$. The fitted profiles are centered
              in velocity on the spectroscopic redshifts listed in Table~\ref{table:emission-lines}.}
     \label{figure:other-emission-lines}%
 \end{figure}

\begin{figure*}
   \centering
\includegraphics[angle=90, width=1.0\textwidth]{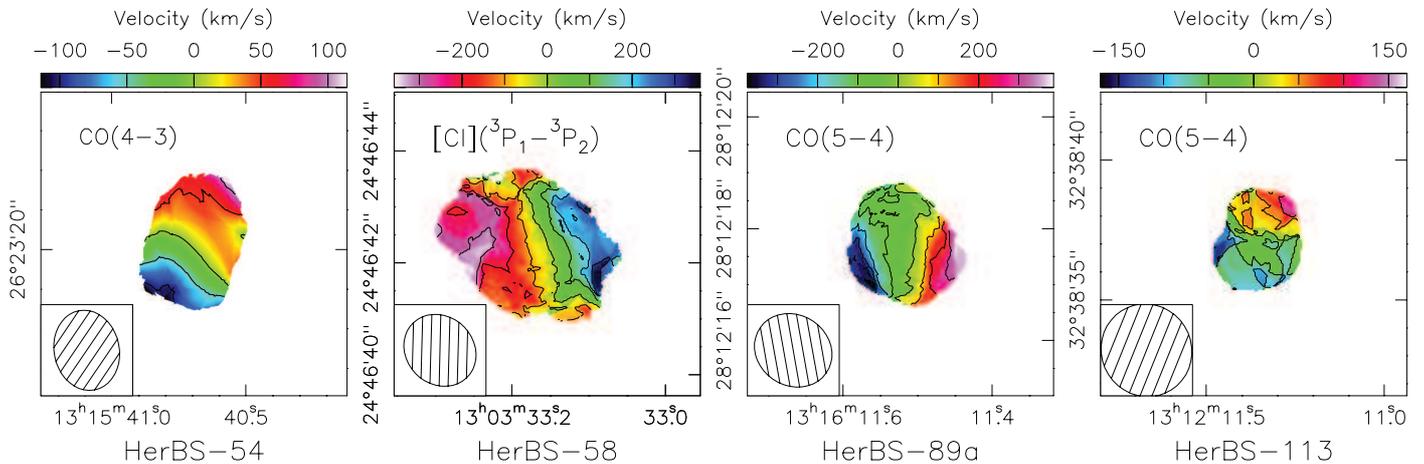}
   \caption{Velocity field maps of HerBS-54, HerBS-58, HerBS-89a, and HerBS-113. The maps were obtained for the 
   emission lines of $^{12}$CO\,(4-3), [C{\small I}]\,($^3$P$_1$-$^3$P$_2$), and $^{12}$CO\,(5-4) 
   above thresholds of 30\%, 20\%, 10\%, and 20\% of 
   the peak emission in the zeroth moment map for HerBS-54, HerBS-58, HerBS-89a, and HerBS-113, respectively. Contours are in units 
   of 50\kms\ for HerBS-54 and HerBS-113, and of 100\kms\ for HerBS-58 and HerBS-89a. The synthesized beams are shown in the 
   lower left corner of each panel.}
   \label{figure:moment-one}
\end{figure*} 

   \begin{figure*}
   \centering
\includegraphics[width=0.65\textwidth]{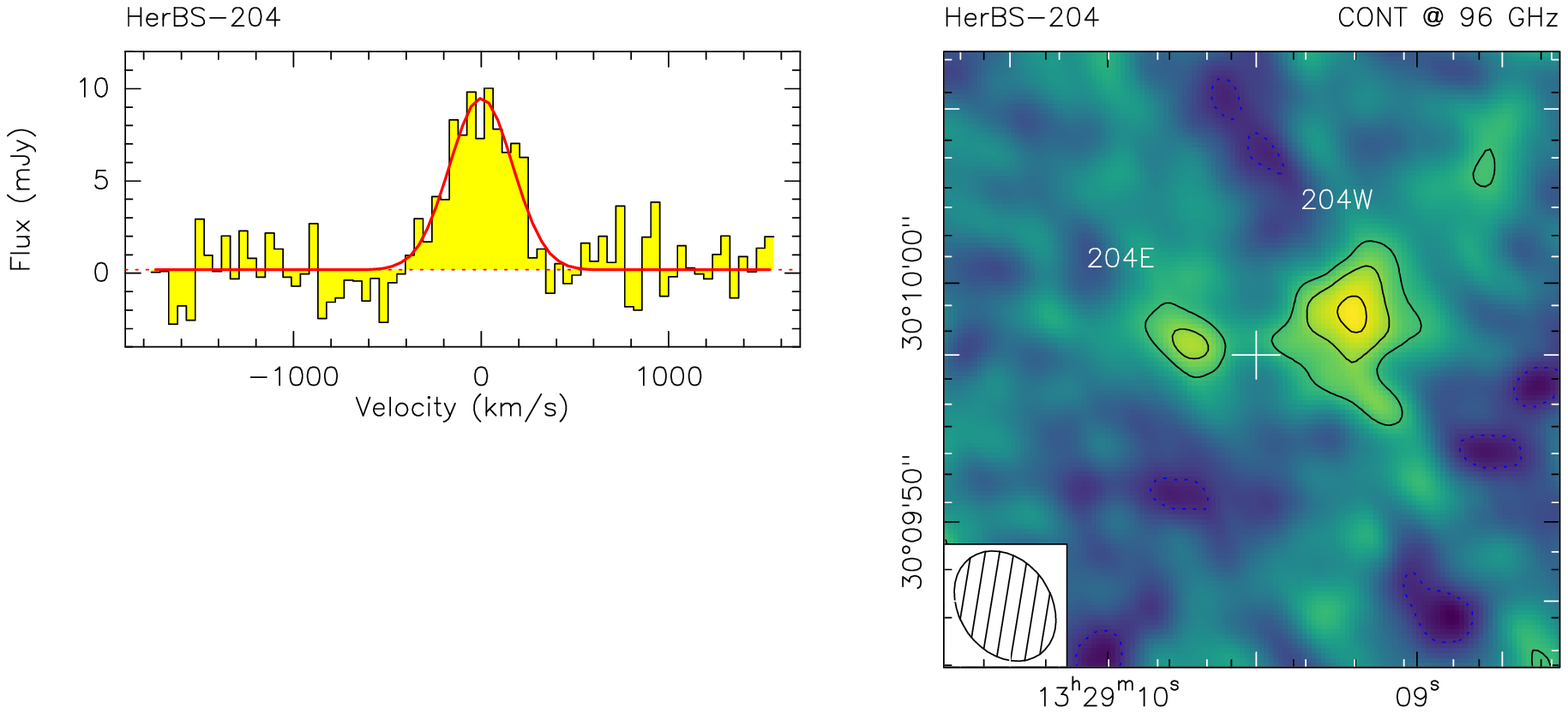}
\includegraphics[width=0.55\textwidth]{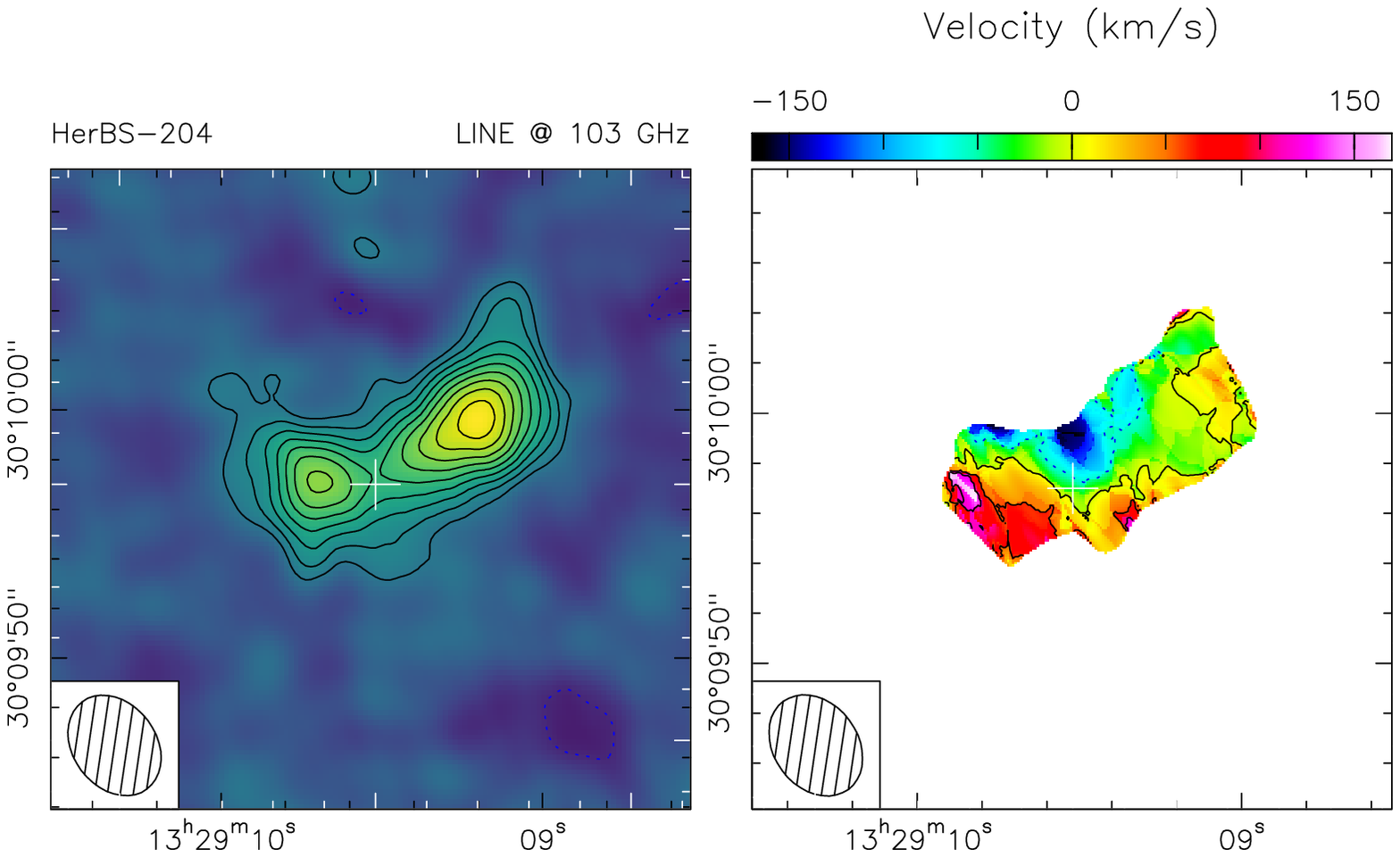}
   \caption{{\it (Top panel)} Integrated spectrum (left) and continuum image (right) for HerBS-204. 
   The continuum contours start at $2\sigma$ and are spaced in steps of $1\sigma$ $\rm = 31\,\mu$Jy\,beam$^{-1}$.
   Fits to the continuum and the integrated emission line profile are shown as dotted and solid red lines, respectively.
   {\it (Bottom panel)} Velocity integrated image of the emission line at 102.584\,GHz (left) and velocity 
   field map of HerBS-204. The velocity map was obtained for the emission line above a threshold of 20\% of 
   the peak emission in the velocity integrated image. Line contours start at $2\sigma$ and are shown in steps 
   of $1\sigma$ $\rm = 0.15 \, Jy \, km s^{-1}$; velocity contours are in units of 50\,km\,s$^{-1}$. 
   The synthesized beams are shown in the lower left corners.} 
   \label{figure:HerBS-204}%
    \end{figure*}

\section{Results}
\label{section:results}

In the 13 fields observed in the Pilot Program, we detected 12  individual sources 
both in the continuum and in at least two lines at 3 and 2 mm. We searched for sources in each field up to a distance of 1.5$\times$ the half width at half maximum of the 3 and 2 mm primary beams. Sensitivity was the main limitation to searching beyond this area.  A source is claimed to be detected if it is detected with at least 5$\sigma$ in two emission lines, and if the positions of the peaks of the corresponding velocity integrated line maps are coincident within the relative astrometric uncertainties of the data. 
Figures~\ref{figure:spectra1} to \ref{figure:spectra4} present a representative 2 mm continuum image and the spectra 
of the two strongest emission lines for each of the sources that were detected in two or more lines, and for which
reliable redshifts were derived. 

The fields in Figs.~\ref{figure:spectra1}, \ref{figure:spectra2}, 
and \ref{figure:spectra3} all show sources that lack companions. In two of the fields a second source 
is detected within the primary beam. 
In the case of HerBS-43 the second galaxy, which is seen in both the continuum and CO emission lines, 
is at a different redshift; in the case of HerBS-89 a nearby galaxy is only detected in the 2 mm continuum, and
is probably unrelated (see Sect.\ref{section:individual-sources}). 
Figure~\ref{figure:spectra4} displays the sources HerBS-70 and HerBS-95, 
which are two binary galaxies with separations of $\sim$16-17$^{\prime\prime}$.
Additional emission lines detected in some of these sources are displayed in Fig.~\ref{figure:other-emission-lines}. 
Finally, two sources, HerBS-173 and HerBS-204, were detected in the 3 mm continuum 
with low signal-to-noise ratio, and in the case of HerBS-204 in a strong emission line (see Fig.~\ref{figure:HerBS-204}); 
neither of these sources was observed at 2 mm. 

The coordinates of the 2 mm continuum emission peaks are 
given in Table~\ref{tab:flux-densities} together with information on the continuum flux densities of the sources. 
Table~\ref{table:emission-lines} lists all the emission lines that were detected with their line fluxes and widths 
and the derived spectroscopic redshifts ($z_{\rm spec}$).  In total, taking into account the companions, 
we provide continuum fluxes for 18 sources (Table~\ref{tab:flux-densities}) and derive spectroscopic 
redshifts for 14 of them (Table~\ref{table:emission-lines}).

\subsection{Individual sources}
\label{section:individual-sources}
In this section we provide a detailed description for each source that was observed in the Pilot Program.

\begin{itemize}
       \item HerBS-34 is a strong continuum source with a flux density of $\rm S_{159\,GHz} = 3.75\pm0.04\,mJy$ 
            that is resolved by the $\sim$1$\farcs3$ beam with an estimated size of $0\farcs7 \pm 0\farcs1$
            (Fig.~\ref{figure:spectra1} upper panel). The
            $\rm ^{12}$CO\,(3-2) and double-peaked (5-4) emission lines are strong, with widths of 
            $\rm \sim$360\,km\,s$^{-1}$, showing an extended structure with an estimated size that is comparable 
            to that of the continuum emission. The derived redshift is $z_{\rm spec} = 2.6637$.
     \item The field of view of HerBS-43 reveals two sources located symmetrically with respect to 
            the phase tracking center that are separated by $\sim$7$\farcs7$ (Fig.~\ref{figure:spectra1} middle panel): 
            \begin{itemize} 
                \item The stronger source (HerBS-43a) is located to the west and has a flux density 
                      of $\rm S_{149\,GHz} =2.6\pm0.3\,mJy$. The $\rm ^{12}CO$\,(4-3) and (5-4) are 
                      very broad ($\rm FWHM \sim$1070\,km\,s$^{-1}$) and double-peaked. The derived redshift is 
                      $z_{\rm spec} = 3.2121$. The emission in both the line and continuum 
                      is unresolved within the 3 mm $1\farcs8 \times 1\farcs6$ beam. 
                \item The second source to the east (HerBS-43b) is weaker, with 
                      $\rm S_{149 \, GHz} = 1.7 \pm 0.3 \, mJy$, and is also unresolved. The
                      profile of the $\rm ^{12}CO$\,(4-3), (5-4), and (6-5) emission lines 
                      (the last line is shown in Fig.~\ref{figure:other-emission-lines}) is  
                      distinct from that of HerBS-43a, also double-peaked but slightly narrower ($\rm 800 \, km \, s^{-1}$). 
                      The derived redshift is different from HerBS-43a, with $z_{\rm spec} = 4.0543$. 
                      The galaxies HerBS-43a and b are hence unrelated. 
            \end{itemize}
      \item HerBS-44 displays a well-defined 2 mm continuum          with $\rm S_{149 \, GHz} = 1.7 \pm 0.3 \, mJy$, and       strong broad (\rm $\sim$520\,km\,s$^{-1}$) 
            emission lines of $\rm ^{12}$CO\,(3-2) and (5-4), yielding a redshift of 
            $z_{\rm spec} = 2.9268$ (Fig.~\ref{figure:spectra1} bottom panel). The source is resolved in 
            the $^{12}$CO\,(3-2) emission line with an estimated size of $1\farcs1 \pm 0\farcs2$.    
      \item HerBS-54 shows a weak 2 mm continuum with a flux density of $\rm S_{134 \, GHz} = 1.6 \pm 0.2 \,mJy$ (Fig.~\ref{figure:spectra2} top panel). 
            The $\rm ^{12}$CO\,(3-2) and (4-3) emission lines are very broad ($\rm \sim$1020\,km\,s$^{-1}$), 
            and the spectroscopic redshift is $z_{\rm spec} = 2.4417$. 
            The source is resolved in the $\rm ^{12}$CO\,(3-2) line with a size of $1\farcs4 \pm 0\farcs3$ 
            and a velocity gradient along a position angle of $\sim$150\,deg (Fig.~\ref{figure:moment-one} left panel). 
      \item HerBS-58 shows slightly extended continuum emission       with a flux density of 
            $\rm S_{159\,GHz} = 1.71\pm0.05\,mJy$. The emission lines of $\rm ^{12}CO$\,(3-2) and 
            $\rm [C{\small I}]\,(^3P_1$-$\rm ^3P_0)$ (Fig.~\ref{figure:spectra2} middle panel) 
            are clearly detected. The $\rm ^{12}CO$\,(4-3) line is also detected, but shows a lower signal-to-noise ratio (see Table\,\ref{table:emission-lines});  its spectral profile is not shown here, as the 
            line is located at the intersection of two correlator basebands.
            The lines are double-peaked and very broad with widths of $\rm \sim$970\,km\,s$^{-1}$. The redshift 
            of HerBS-58 is $z_{\rm spec} = 2.0842$ (Fig.~\ref{figure:spectra2} middle panel). The line emission
            is resolved with a size of $1\farcs6$, and shows a hint of a velocity gradient 
            in the east--west direction (Fig.~\ref{figure:moment-one} middle panel). The possibility of a binary system cannot be disregarded for this particular object.
      \item HerBS-70 is a binary system in which both sources are at the same redshift and 
            have a large separation of $\sim$16$\farcs5$ (Fig.~\ref{figure:spectra4} upper panel). The eastern source (HerBS-70E)
            has a 2 mm continuum flux density of $\rm S_{159\,GHz}=0.94\pm0.04\,mJy$ and is resolved with 
            a size of 0.5$''$. The source to the west (HerBS-70W) is weaker, with a 
            primary beam corrected flux density of $\rm S_{159\,GHz} = 0.18\pm0.06\, mJy$. 
            The source HerBS-70E has strong double-peaked asymmetrical emission lines of $\rm ^{12}CO$\,(3-2) 
            and (4-3) with widths of $\rm \sim$770\,km\,s$^{-1}$. In contrast, HerBS-70W displays significantly narrower
            ($\sim$140\,km\,s$^{-1}$) single-peaked emission lines, suggesting a face-on inclination. 
            Both sources are at the same redshift with $z_{\rm spec} = 2.31$, implying a projected
            distance of $\sim$140\,kpc between HerBS-70E and HerBS-70W. 
      \item HerBS-79 shows a weak 2 mm continuum that is barely detected at the sensitivity of 
            the current observations ($\rm S_{149\,GHz}=0.8\pm0.3\,mJy$). The very broad (\rm $\sim$870\,km\,s$^{-1}$) 
            emission lines of $\rm ^{12}CO$\,(3-2) and (4-3) display similar double-peaked profiles, with the 
            red component being about three times more intense than the blue one (Fig.~\ref{figure:spectra2} bottom panel). 
            The derived redshift is $z_{\rm spec} = 2.0782$. The source is 
            resolved in the $\rm ^{12}CO$\,(3-2) emission line with an estimated size of $1\farcs1 \pm 0\farcs2$.  
      \item HerBS-89 is a system composed of two objects, of which HerBS-89a is the strongest 2 mm continuum 
            source in the sample, with a flux density of $\rm S_{159\,GHz}=4.56\pm0.05\,mJy$ (Fig.~\ref{figure:spectra3} top panel). 
            The 2 mm continuum emission is resolved by the $1\farcs3 \times 1\farcs2$ beam, with an 
            extension of 0$\farcs9 \pm 0\farcs1$. The $\rm ^{12}$CO\,(3-2) and (5-4) emission lines are 
            also the broadest in the sample ($\rm \sim$1080\,km\,s$^{-1}$), displaying a double-peaked 
            profile. The redshift of HerBS-89a is $z_{\rm spec} = 2.9497$. The CO line emission is also extended 
            and displays an east--west velocity gradient (Fig.~\ref{figure:moment-one} right panel). 
            Follow-up observations with NOEMA at higher frequency, 
            with an angular resolution of $0\farcs3$, reveal a 
            nearly complete Einstein ring in the 1 mm dust continuum and the
            $\rm ^{12}CO$\,(9-8) and para-$\rm H_2O\,(2_{02}$-$1_{11})$ line emission, showing that HerBS-89a is 
            gravitationally lensed (Berta et al. in preparation). To the east of HerBS-89a is a weak 
            unresolved source (HerBS-89b) with a flux density of $\rm S_{159 \, GHz} = 0.24 \pm 0.05 \, mJy$. Although there is no corresponding source in the SDSS catalogue at that position, its authenticity is confirmed by the higher frequency measurements (Berta et al. in preparation). Further observations 
            are needed to constrain the properties of HerBS-89b. 
      \item HerBS-95 is another binary system in which both sources are at the same redshift 
            with a separation of $\sim$16\farcs4 (Fig.~\ref{figure:spectra4} lower panel). 
            The eastern source (HerBS-95E) exhibits a continuum flux density at 2 mm of $\rm S_{159\,GHz}=1.52\pm0.04\,mJy$ 
            and a size of $0\farcs5$, whereas the western source (HerBS-95W) has a primary beam corrected flux density of 
            $\rm S_{159\, GHz}=2.28\pm0.08\,mJy$. Both sources show strong emission lines of $\rm ^{12}CO$\,(3-2) and (5-4), 
            with linewidths of 870 and $\rm 540 \, km \, s^{-1}$ for HerBS-95E and W, respectively. The lines are at nearly the same 
            frequencies, indicating that both galaxies are at a redshift of $z_{\rm spec} = 2.97$. 
            At this redshift the projected distance between the two galaxies is $\sim$140\,kpc. 
      \item HerBS-113 has a weak 2 mm continuum with a 
            flux density of  $\rm S_{149 \, GHz} = 1.4 \pm 0.3 \, mJy$ (Fig.~\ref{figure:spectra3} middle panel). 
            The $\rm ^{12}CO$\,(3-2) 
            and (5-4) emission lines are well detected, displaying broad profiles with widths 
            of $\rm \sim$900\,km\,s$^{-1}$. Both emission lines are resolved with an elongation along a position 
            angle of 20\,$\deg$ over a region of $\sim$1$\farcs2$ (Fig.~\ref{figure:moment-one}). 
            The derived redshift is $z_{\rm spec} = 2.7870$. 
      \item HerBS-154 is a compact source with a size of $1\farcs2$ in continuum and line emission (Fig.~\ref{figure:spectra3} 
            bottom panel). The source is robustly detected in the continuum with $\rm S_{149 \, GHz} = 1.92\pm 0.04 \, mJy$, and in the lines of $\rm ^{12}CO$\,(6-5) and $\rm H_2O\,(2_{11}$-$2_{02})$, 
            and in $\rm [C{\small I}]\,(^3P_1$-$\rm ^3P_0)$ (shown in Fig.~\ref{figure:other-emission-lines}), 
            although with lower signal-to-noise ratio. The spectral profiles, which are single-peaked with linewidths 
            of $\sim$300\,km\,s$^{-1}$, yield a redshift of $z_{\rm spec}=3.7070$.  
      \item The sources HerBS-173 and HerBS-204 have the weakest 500~$\rm \mu m$ flux densities in the 
            Pilot Program sample (Table~\ref{tab:flux-densities}).  
            \begin{itemize} 
            \item HerBS-173 was tentatively detected in the individual 3 mm sidebands.  
                Stacking these sidebands results in a 3 mm flux density of $\rm S_{100 \, GHz} = 0.22 \pm 0.03 \, mJy$. 
                However, no emission line was detected at 3 mm and no 2 mm observations were performed.
                Hereafter, we adopt the photometric redshift $z_{\rm phot} = 2.38$ (see Table~\ref{table:sources}).\\
                \item In the case of HerBS-204, stacking the line-free part of the 3 mm spectra observed on Aug 6 and 7, 2019, in both 
                sidebands (LSB and USB), reveals a complex source
                with two continuum emission peaks (HerBS-204E and HerBS-204W) separated by $\sim$7$''$ (Fig.~\ref{figure:HerBS-204})  and with fluxes of $\rm S_{96 \, GHz} = 0.10\pm 0.03 \, mJy$ and $\rm S_{96 \, GHz} = 0.13\pm 0.03 \, mJy$, respectively. 
                In addition, a strong emission line is detected at 102.584\,GHz with a linewidth of $\sim$400\,km\,s$^{-1}$ and 
                an integrated line flux of 3.9\,Jy\,km\,s$^{-1}$ (Table~\ref{table:emission-lines}). Like the continuum emission, the 
                line emission is extended with two emission peaks separated by $6\farcs8$ along a position angle of 
                $\sim$18\,$\deg$. Both the continuum and line emission 
                peaks show excellent spatial coincidence. Based on the photometric redshift of $z_{\rm phot}=3.61$, this emission 
                line could correspond either to the $^{12}$CO\,(4-3) transition, in which case the source would be at a 
                spectroscopic redshift of $z_{\rm spec} = 3.49$, or to the $^{12}$CO\,(3-2) transition, 
                in which case $z_{\rm spec} = 2.37$. The higher value 
                ($z_{\rm spec} = 3.49$) would imply a dust temperature of 40\,K, which is at the high end of the 
                dust temperatures found for all the other sources of the Pilot Program. This suggests 
                the value of $z_{\rm spec} = 2.37$, for which the estimated dust temperature is 29\,K,
                is the more likely redshift (see Table~\ref{tab:luminosities} and footnote).
                However, further observations are needed to detect a second CO transition and 
                derive a reliable spectroscopic redshift for this source. Based on the photometric redshift 
                and the potential range in the spectroscopic redshift, the projected separation between the 
                two emission peaks corresponds to a linear distance of $\rm \sim$60\, kpc, suggesting that 
                HerBS-204 is a merging system or a gravitationally lensed galaxy rather than an edge-on disk \cite[cf.][]{Emonts2018}. 
                 \end{itemize}
   \end{itemize}

  \begin{figure}
   \centering
\includegraphics[width=0.48\textwidth]{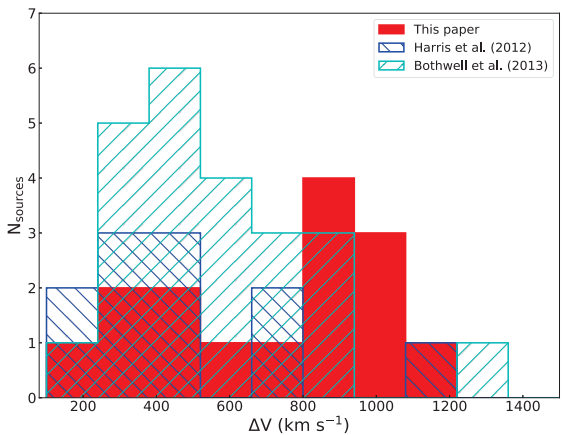}
   \caption{Distribution of the full width at half maximum (FWHM) for the bright {\it Herschel} galaxies detected 
   in $\rm ^{12}CO$ described in this paper (solid red histogram) compared to the FWHM of high-$z$ galaxies reported 
   in \cite{Bothwell2013} (cyan right-hatched) and \cite{Harris2012} (blue left-hatched).}
   \label{figure:line-widths}%
    \end{figure}

\subsection{Spectroscopic redshifts and emission line properties}
\label{section:emission-line-properties}

For all of the above sources (except HerBS-204 and HerBS-173), we detect at least two emission lines, mostly from $^{12}$CO 
ranging from the (3$-$2) to the (6$-$5) transition (Table~\ref{table:emission-lines}). The CO emission lines are all relatively strong,
resulting in signal-to-noise ratios $>5$, providing therefore 
the necessary quality to derive precise and reliable redshifts, as well as significant information 
about properties of the molecular gas such as morphology, dynamics, and physical conditions.
In addition to the CO emission lines, the atomic carbon fine-structure line $\rm [C{\small I}]\,(^3P_1$-$^3$P$_0)$ 
is detected in three sources, HerBS-58 (Fig.~\ref{figure:spectra2}), and HerBS-70E and HerBS-154
(Fig.~\ref{figure:other-emission-lines}). For HerBS-154 we detected the para-H$_2$O\,($2_{11}$-$2_{02}$) 
transition (see Fig.~\ref{figure:spectra3}).

\rm It is noteworthy that for the majority of the bright {\it Herschel} galaxies observed in the Pilot Program, 
the widths of the emission lines are large. The derived linewidths are found to be 
150\,km\,s$^{-1} < \Delta V <1100$\,km\,s$^{-1}$ with most (53\%) of the sources having linewidths in excess 
of $\rm 800\,km\,s^{-1}$ (see Table~\ref{table:emission-lines} and Fig.~\ref{figure:line-widths} showing 
the distribution of the CO emission linewidths, and further discussion in Sect. \ref{section:linewidths}).

\subsection{Continuum}
\label{section:continuum}

In addition to the emission lines, the continuum flux densities of the sources have been extracted from up to eight 
available polarization-averaged 7.744\,GHz wide sidebands, centered, depending on receiver configuration, on the following frequencies: 88.3, 96.0, 
103.7, 111.5, 133.5, 143.2, 149.0, and 158.6\,GHz (see Tables~\ref{table:Observing_Log} and \ref{tab:flux-densities}). 
All the sources in the Pilot Program detected in at least two emission lines are also detected in the continuum in at least four sidebands, 
with three in all eight sidebands (one of which, HerBS-95, being double). The NOEMA continuum flux densities together with the {\it Herschel} and
SCUBA-2 data are displayed in Fig.~\ref{figure:sed_fit} for HerBS-58 and HerBS-89a.  
The NOEMA continuum measurements are summarized in Table~\ref{tab:flux-densities}, 
where the quoted flux density uncertainties account for both the noise level in the maps 
and the uncertainty in the absolute flux calibration scale (see Section \ref{section:results}). 
In Table~\ref{tab:flux-densities}, upper limits are given for sources where the continuum 
is detected with a signal-to-noise ratio $<4$.

\begin{table*}[!ht]
\tiny 
\centering
\caption{Observed continuum positions and flux densities}
\begin{tabular}{lcccccccccc}
\hline
\hline
Source & RA & Dec. & \multicolumn{8}{c}{S$_\nu$ $($mJy$)$}  \\
        & \multicolumn{2}{c}{(J2000)} & 158.6 GHz &     149.0 GHz &     143.2 GHz &   133.5 GHz &     111.5 GHz &     103.7 GHz &     96.0 GHz &      88.3 GHz \\
\hline
HerBS-34  & 13:34:13.87 & 26:04:57.5 & 3.75$\pm$0.04  &   -- &  2.61$\pm$0.03  &          --            &     0.65$\pm$0.12  &        0.79$\pm$0.06  &        0.47$\pm$0.09  &      0.37$\pm$0.06   \\ 
HerBS-43a &  13:24:18.79 & 32:07:54.4 &           --       & 2.6$\pm$0.3 &          --             &     1.6$\pm$0.2  &  0.63$\pm$0.08  &        0.49$\pm$0.06  &      0.29$\pm$0.04  &        0.12$\pm$0.06   \\ 
HerBS-43b &  13:24:19.24 & 32:07:49.2 &           --       & 1.7$\pm$0.3 &          --             &     1.2$\pm$0.2  &  0.74$\pm$0.08  &        0.29$\pm$0.06  &      0.31$\pm$0.04  &        0.25$\pm$0.06   \\ 
HerBS-44  &   13:32:55.85 & 34:22:08.4 &    --     & 1.7$\pm$0.3 &         --          & 1.5$\pm$0.2  &      --        & 0.35$\pm$0.07  &            --         & 0.22$\pm$0.06    \\
HerBS-54  &      13:15:40.72 & 26:23:19.6 & --     & 1.7$\pm$0.3 &         --              &     1.6$\pm$0.2  &      --            &     0.45$\pm$0.07  &           --           &     0.23$\pm$0.06   \\  
HerBS-58  & 13:03:33.17 & 24:46:42.3 & 1.71$\pm$0.05  & 1.0$\pm$0.3 &   1.19$\pm$0.04  &      0.7$\pm$0.2  &  0.37$\pm$0.07  &        0.25$\pm$0.07  &        0.22$\pm$0.05  &      0.18$\pm$0.06   \\ 
HerBS-70E  & 13:01:40.33 & 29:29:16.2 & 0.94$\pm$0.04  & 0.9$\pm$0.3 &  0.60$\pm$0.06  &      0.6$\pm$0.2  &      --            &     0.25$\pm$0.07 &      --           &      0.22$\pm$0.06   \\ 
HerBS-70W  & 13:01:39.31 & 29:29:25.2 & 0.18$\pm$0.06 &     <0.8     &  0.16$\pm$0.06 &           <0.7          &         --            &         <0.4     &       --            &             <0.2        \\ 
HerBS-79   &  13:14:34:08 & 33:52:20.1 &     --    & 0.8$\pm$0.3 &         --          &     <0.7          &     0.30$\pm$0.06  &        0.22$\pm$0.06  &      0.16$\pm$0.04  &        0.13$\pm$0.05    \\ 
HerBS-89a   & 13:16:11.52 & 28:12:17.7 & 4.56$\pm$0.05  & 3.4$\pm$0.3 & 3.02$\pm$0.04  &      2.2$\pm$0.2  &  1.10$\pm$0.08  &        0.83$\pm$0.05  &        0.56$\pm$0.06  &      0.44$\pm$0.06   \\ 
HerBS-89b   & 13:16:11.93 & 28:12:16.7 & 0.24$\pm$0.05  & <0.1 &        --  &      --  &   --  &   --  &   --  &   --      \\ 
HerBS-95E  & 13:43:42.73 & 26:39:18.0 & 1.52$\pm$0.04  & 1.1$\pm$0.3 &  1.07$\pm$0.03  &      0.6$\pm$0.2  &  0.30$\pm$0.06  &        0.16$\pm$0.07  &        0.13$\pm$0.04  &      0.10$\pm$0.06   \\ 
HerBS-95W  &  13:43:41.55 & 26:39:22.7 & 2.28$\pm$0.08  & 2.1$\pm$0.4 & 1.34$\pm$0.06  &      0.7$\pm$0.3  &  0.27$\pm$0.09 & 0.18$\pm$0.08  &        0.12$\pm$0.05 &       0.14$\pm$0.07   \\ 
HerBS-113  &  13:12:11.35 & 32:38:37.8 &        --                 & 1.4$\pm$0.3 &          --             &     0.8$\pm$0.2  &  0.75$\pm$0.06  &                 --        &     0.46$\pm$0.04  &        --      \\ 
HerBS-154  & 13:22:58.11 & 32:50:51.7 & 1.92$\pm$0.04  & 1.6$\pm$0.3 &  1.39$\pm$0.03  &      0.9$\pm$0.2  &      --            &     0.25$\pm$0.06 &      --           &      0.13$\pm$0.06   \\ 
HerBS-173  & 13:18:04.15 & 32:50:15.9  & --      &      --                &          --            &         --            &     0.19$\pm$0.16  &        0.29$\pm$0.12  &      0.22$\pm$0.10  &        0.33$\pm$0.12   \\
HerBS-204E & 13:29:09.74& 30:09:57.5 &  --                 &    --                &          --            &         --            &       <0.1        &     <0.1   &       0.10$\pm$0.03           &             <0.1        \\
HerBS-204W & 13:29:09.21& 30:09:58.7 &  --                 &    --                &          --            &         --            &       <0.1        &     <0.1   &       0.13$\pm$0.03 &               <0.1        \\
\hline
\end{tabular}
\tablefoot{Positions are derived from the 2 mm continuum peaks, with the exception of HerBS-173 and HerBS-204, whose positions 
are derived from the stacked 3 mm continuum peaks (see Sect.~\ref{section:individual-sources} for further details). 
The width of each of the sidebands is 7.744~GHz, and their frequency ranges are provided in Table~\ref{table:Observing_Log}. 
The considerably longer integration times in Frequency Setting\,4 resulted in better sensitivities in the corresponding 2 mm sidebands. 
See Section~\ref{section:individual-sources} for the continuum flux densities of the sources HerBS-204 and HerBS-173. 
For HerBS-70W and HerBS-95W, the flux densities and upper limits were corrected for primary beam attenuation.}
\label{tab:flux-densities}
\end{table*}
\normalsize 

\begin{table*}[!htbp]
\tiny
\caption{Summary of emission line properties and spectroscopic redshifts}
\begin{tabular}{lcccccccc}
\hline\hline
Source & $z_{\rm spec}$ & $\Delta V$\,(km \, s$^{-1})$ & \multicolumn{5}{c}{Line Flux $\rm (Jy \, km \, s^{-1})$}  \\
       &  & & $^{12}$CO\,(3-2) & $^{12}$CO\,(4-3) &  $^{12}$CO\,(5-4)  & $^{12}$CO\,(6-5)  & $\rm [C{\small I}]\,(^3P_1$-$\rm ^3P_0)$ 
       &  H$_2$O($2_{11}$-$2_{02}$)\\
\hline
HerBS-34  & 2.6637\,(2) &  330$\pm$10  & 2.8$\pm$0.4 &     --      &  8.4$\pm$0.8 &     --      &     --      & --          \\ 
HerBS-43a & 3.2121\,(1) & 1070$\pm$90  &    --       & 5.5$\pm$0.8 &  6.7$\pm$0.8 &     --      &     --      &  --         \\ 
HerBS-43b & 4.0543\,(7) &  800$\pm$50  &    --       & 1.7$\pm$0.3 &  1.5$\pm$0.3 & 4.8$\pm$0.9 &   <0.7      & <2.9        \\ 
HerBS-44  & 2.9268\,(2) &  520$\pm$50  & 4.9$\pm$0.5 &     --      & 12.5$\pm$1.2 &     --      &     --      & --          \\ 
HerBS-54  & 2.4417\,(3) & 1020$\pm$190 & 4.3$\pm$0.4 & 8.5$\pm$0.8 &     --       &     --      &     --      & --          \\ 
HerBS-58  & 2.0842\,(1) &  970$\pm$50  & 5.3$\pm$1.0 & 5.2$\pm$1.5 &     --       &     --      & 4.7$\pm$0.5 & --          \\
HerBS-70E & 2.3077\,(4) &  770$\pm$50  & 1.8$\pm$0.5 & 3.4$\pm$0.3 &     --       &     --      & 3.5$\pm$0.7 & --          \\ 
HerBS-70W & 2.3115\,(1) &  140$\pm$20  & 1.7$\pm$0.3 & 2.0$\pm$0.3 &     --       &     --      &       <2.9  & --          \\
HerBS-79  & 2.0782\,(8) &  870$\pm$70  & 4.1$\pm$0.8 & 5.5$\pm$0.5 &     --       &     --      &     --      & --          \\ 
HerBS-89a  & 2.9497\,(1) & 1080$\pm$60  & 4.0$\pm$0.6 &     --      &  8.4$\pm$0.8 &     --      &     --     & --          \\ 
HerBS-95E & 2.9718\,(3) &  870$\pm$50  & 1.0$\pm$0.1 &     --      &  3.6$\pm$0.3 &     --      &     --      & --          \\ 
HerBS-95W & 2.9729\,(2) &  540$\pm$30  & 2.4$\pm$0.4 &     --      &  3.5$\pm$0.3 &     --      &     --      & --          \\
HerBS-113 & 2.7870\,(8) &  900$\pm$200 & 6.1$\pm$1.2 &     --      & 13.5$\pm$1.4 &     --      &       <2.7  & --          \\ 
HerBS-154 & 3.7070\,(5) &  310$\pm$40  &     --      &     --      &     --       & 7.6$\pm$0.7 & 1.3$\pm$0.4 & 1.5$\pm$0.3 \\
HerBS-204 &      --     &  400$\pm$80  & --          &     --      &     --       &     --      &     --      & --          \\
\hline
\end{tabular}
 \tablefoot{The uncertainties in the spectroscopic 
 redshifts, $z_{\rm spec}$, given in brackets, correspond to the last decimal derived from Gaussian fits to 
 the line profiles. $\Delta V$ values are the mean linewidths (FWHM) weighted by the peak intensities of the detected CO transitions. The linewidths of double-peaked profiles were estimated as $\Delta V = \Delta w + \Delta s$ by fitting two Gaussians of identical width $\Delta w$ and separation $\Delta s$. Linewidths and their uncertainties are rounded to the closest multiples
 of 10\,km\,s$^{-1}$. For the sources that remain undetected in the [C{\small I}]\,($^3$P$_1$-$^3$P$_0$) 
 and H$_2$O\,($2_{11}$-2$_{02}$) transitions, we provide upper limits to the line fluxes. 
 The upper limits are based on 3$\sigma_v\sqrt{\Delta V\Delta v}$, 
 where $\Delta v$ and $\sigma_v$ are the width in velocity and RMS noise of a spectral channel, respectively.
 In the case of HerBS-70W and HerBS-95W, the line fluxes and the upper limits were corrected for the primary 
 beam attenuation. See Sect.~\ref{section:individual-sources} for a discussion of the possible 
 identification of the emission line detected in HerBS-204.}
   \label{table:emission-lines}
\end{table*}

\section{Discussion} 
\label{section:discussion}

In this section we describe the derived properties of the high-$z$ bright {\it Herschel} galaxies studied in the Pilot Program. 

\subsection{Spectroscopic redshifts}
\label{section:spectroscopic-redshifts}

\begin{figure}
   \centering
\includegraphics[width=0.48\textwidth]{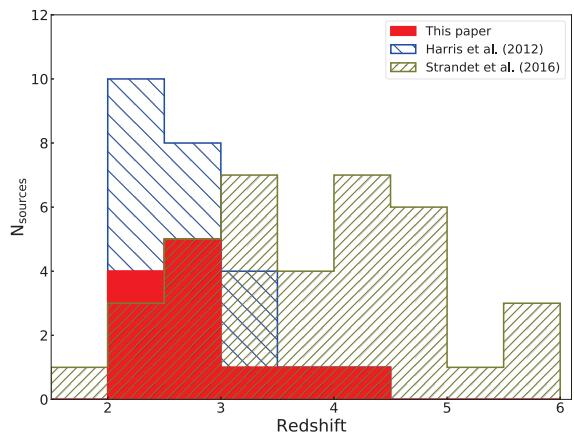}
   \caption{Spectroscopic redshift distribution for the 12 bright {\it Herschel} H-ATLAS galaxies of the Pilot Program 
   sample detected in at least two emission lines red filled histogram, see Table\,\ref{table:emission-lines}). Shown are also the 12 H-ATLAS galaxies with reliable redshifts from \cite{Harris2012} added to the Pilot 
   Program sample (blue left-hatched histogram) and  the redshift distribution 
   of the 38 SPT-selected galaxies from \cite{Strandet2016} (green right-hatched
histogram).}
   \label{figure:spectroscopic-redshifts}%
    \end{figure}

The availability of at least two emission lines in the 3 and 2 mm spectral bands allowed us to derive precise redshifts for 85\% of the bright high-$z$ {\it Herschel} galaxies studied here. 
The Pilot Program has demonstrated that 
using the new correlator on NOEMA, unbiased redshift surveys can be performed efficiently using on average 100 minutes of 
telescope time per source, including overheads. The derived spectroscopic redshifts, $z_{\rm spec}$, for 
the 14 galaxies (including the binary sources) in which two or more emission lines are detected are listed 
in Table~\ref{table:emission-lines}. The redshift distribution of the Pilot Program sample  
is displayed in Fig.~\ref{figure:spectroscopic-redshifts}. The redshifts are found to lie between $2.08<z_{\rm spec}<4.05$, 
with a median redshift of $z=2.86 \pm 0.56$ 
and a tail in the distribution to $z>3$.
In Fig.\,\ref{figure:spectroscopic-redshifts}, we added 
the redshifts of the 12 H-ATLAS galaxies that were studied by \cite{Harris2012}; these sources, which are peaking at 350\,$\rm \mu m$, like the ones studied here, show 
a similar distribution to the Pilot Program sample, albeit with a slightly lower median 
redshift of $z=2.47\pm0.11$. 

Considering the galaxies with redshifts detected by the SPT \citep{Strandet2016, Weiss2013}, we find that the 
redshift distributions of the H-ATLAS and SPT-selected galaxies are clearly different (Fig.\,\ref{figure:spectroscopic-redshifts}). 
The SPT galaxies show a flat distribution between $z=2.5$ and $z=5.0$, 
the major fraction of the sample being at $z>3$ with a median redshift of $z=3.9\pm0.4$. 
The SPT galaxies were selected from a survey performed at a longer wavelength than H-ATLAS; even though the spectroscopic redshift survey of 
the H-ATLAS galaxies is still not complete, the difference in redshifts between the SPT and H-ATLAS 
selected galaxies is significant, and is consistent with expectations for the selected
wavelengths of the 
surveys \citep[see, e.g.,][and references therein]{Strandet2016, Bethermin2015}. The systematic study of the galaxies 
selected from the {\it Herschel} and SPT surveys thus offers the opportunity to gather critical complementary 
information on galaxy populations at different epochs of cosmic evolution, with {\it Herschel}-selected sources probing the
peak of star formation activity around $2<z<3$, while the SPT-selected galaxies provide crucial information 
on star formation at earlier epochs.

\subsection{Comparison to photometric redshifts}
\label{section:photometric-redshifts}

The spectroscopically derived redshifts significantly differ in many cases from the estimates based on the available 
photometric data. Deriving redshifts using submillimeter spectral energy distributions (SEDs) of galaxies with 
known redshifts and dust temperatures as templates indeed is uncertain \citep[e.g.,][]{Jin2019}. 
This is particularly true when using SPIRE data alone because the 
250, 350, and 500~$\rm \mu m$ bands are close to the peak of the observed SED for $2<z<4$ galaxies. 
\citet{Bakx2018} built an SED template based on the SPIRE and SCUBA-2 data for a sample of bright 
H-ATLAS galaxies with available measurements of $z_{\rm spec}$ and a two-temperature modified blackbody (MBB) model.
This template was then used to derive $z_{\rm phot}$ values for the entire H-ATLAS
sample of the {\it Herschel}-bright galaxies (see revised values in Bakx et al. (in preparation)). 
The $z_{\rm phot}$ values derived for the sources of this Pilot Program are listed in Table~\ref{table:sources}. 

Not counting HerBS-43, the values for $z_{\rm phot}$ are on average 
consistent within 20\% of the $z_{\rm spec}$ value, and for two sources, HerBS-95 and HerBS-113, in agreement within 10\%. The poor accuracy and reliability of redshifts derived from (sub)millimeter continuum photometry is due to the degeneracy between temperature, $\beta$, and redshift,
and to the absence of well-defined features in the SEDs. The derived values of the redshifts based on continuum measurements alone is therefore indicative and, in any case, never precise enough to follow up efficiently with targeted observations of molecular or atomic gas. 

\subsection{Spectral energy distribution: infrared luminosity and dust properties}
\label{section:SED}

\begin{figure*}[!ht]
\centering
\includegraphics[width=0.42\textwidth]{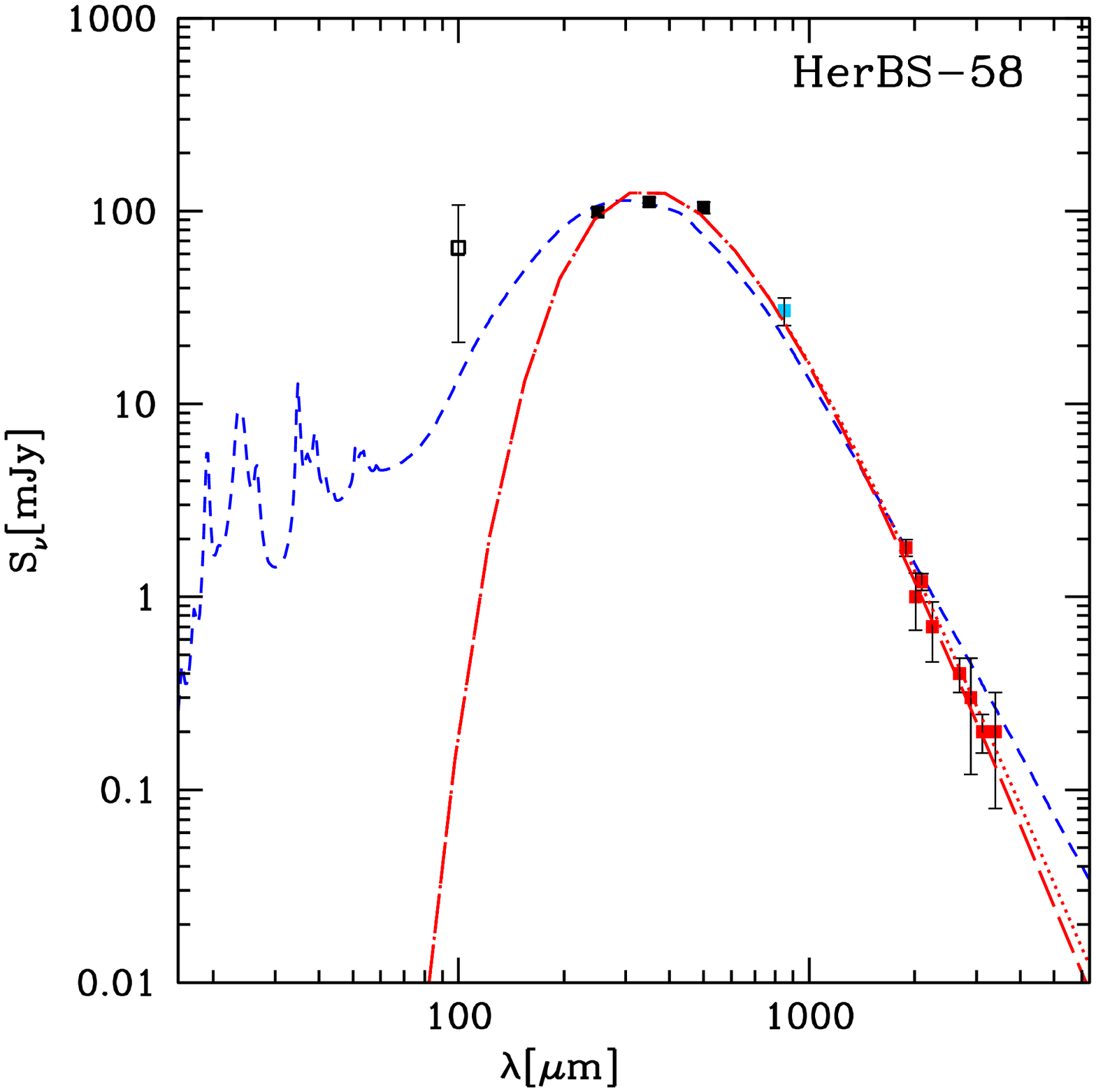}
\includegraphics[width=0.42\textwidth]{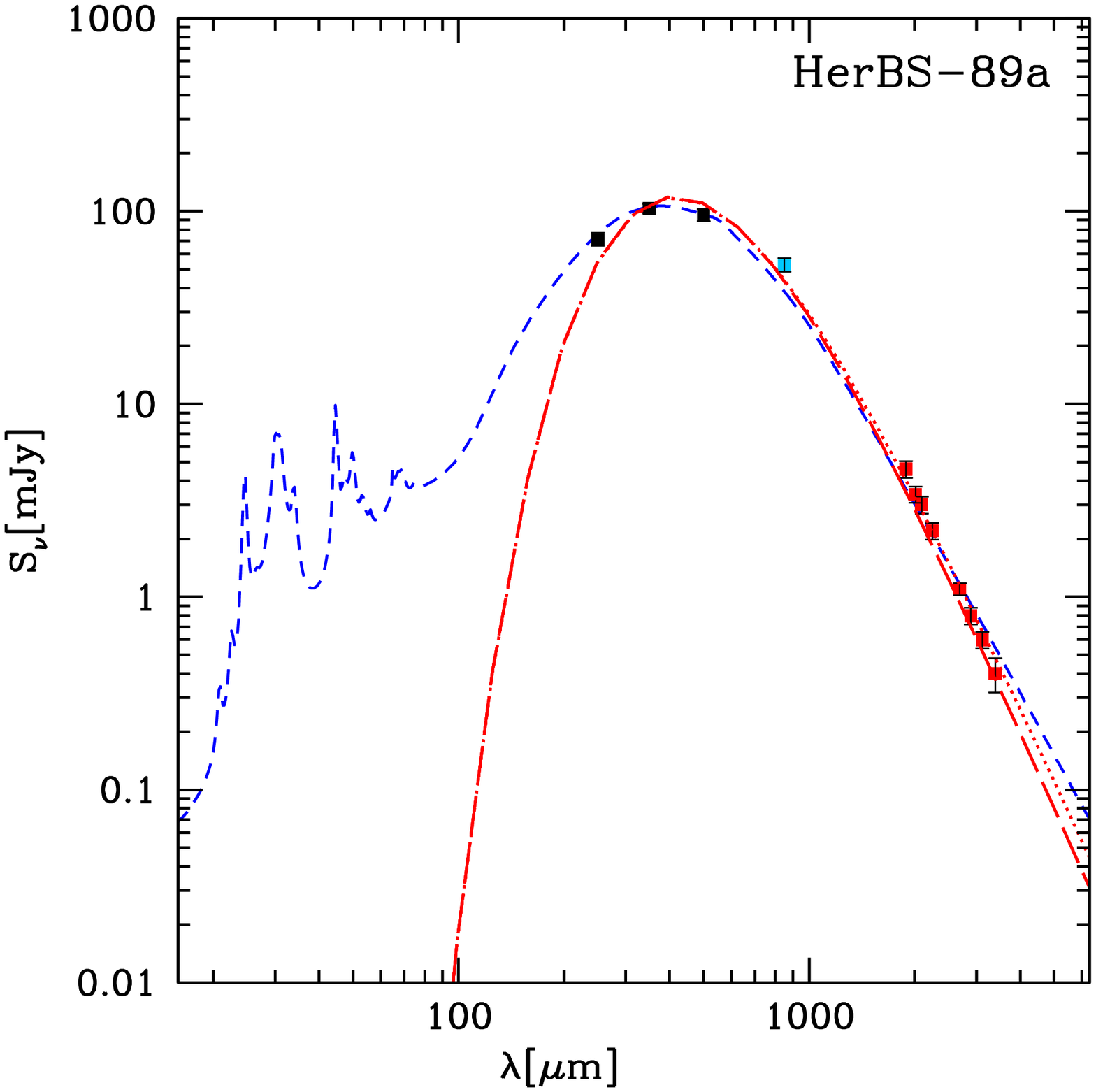}
\caption{Observed  SED of two of the Pilot Program sources, HerBS-58 ($z=2.084$) and 
HerBS-89a ($z=2.950$). The data include SPIRE  \cite[black dots, from][]{Bakx2018}, the revised SCUBA-2 photometry (blue dot,  see 
Bakx et al. \, in preparation), and the 3 and 2 mm continuum flux densities (red dots) extracted 
from the NOEMA data (Table~\ref{tab:flux-densities}). In the case of HerBS-58, the PACS data point, which is
available, is shown as an open square, although it was not used to fit the SED.  The figure also shows the best-fitting MBB model including (red dashed) and not including (red dotted) the effect from the CMB on the dust continuum, 
and the best fit to the DL07 model (blue dashed); see text for details.}
\label{figure:sed_fit}
\end{figure*}

Combining the photometric data from SPIRE \citep{Eales2010} and SCUBA-2 
\citep[][as revised in Bakx et al. in preparation]{Bakx2018} with the 
NOEMA continuum measurements, we assembled the SEDs of all the sources 
observed in the Pilot Program (see below for the cases where the sources are double). 
Although PACS data from the H-ATLAS survey are also available, 
their usefulness is limited as many of the detections are  tentative with signal-to-noise ratio $<$3. 
We have therefore plotted the PACS flux densities, when available, 
on the SEDs, without including them in the SED analysis (see Fig.~\ref{figure:sed_fit}). 
The resulting SEDs cover the observed wavelength range from 250\,$\mu$m to $\sim$3 mm, and include 
sources with a minimum of 7 data points and sources with a maximum of 12. 

In order to derive the infrared luminosities, dust masses, and temperatures of the 
sources, we modeled their SEDs using two different approaches: (i) a single-temperature MBB, following
\cite{berta2016}, and (ii) the  \citet[][hereafter DL07]{DL07} dust models. 

In the first case, the far-infrared SED of a galaxy is modeled as the emergent luminosity 
from a given dust mass M$_\textrm{dust}$: 
\begin{equation}\label{eq:mbb}
L_\nu \sim M_\textrm{dust} \kappa_\nu B_\nu(T_\textrm{dust}) \textrm{,}
\end{equation}
where $B_{\nu}(T_\textrm{dust})$ is the Planck function, $T_\textrm{dust}$ the dust temperature, 
and $\kappa_\nu=\kappa_0\left(\nu/\nu_0\right)^\beta$ the mass absorption coefficient of dust 
at rest frequency $\nu$. 
For $\kappa_\nu$, we adopt the values from \citet{Draine2003}, as 
revised from \citet{li2001}. Ideally, the chosen reference (rest-frame) frequency $\nu_0$ should 
be covered by the observed data. 
We refer to \citet{berta2016} and \citet{Bianchi2013} for a thorough discussion about the proper 
use of $\kappa_\nu$ and assumptions on $\beta$.

For the MBB fit, we limit the observed data to a rest-frame wavelength $\lambda_\textrm{rest}$\,>\,$50\,\mu$m 
in order to avoid biases towards warmer temperatures.
From the MBB modeling, we determine the dust temperature, dust mass, and spectral emissivity index $\beta$ for each source under the assumption that the dust emission is optically thin. The effects of the cosmic microwave background (CMB) discussed in \citet{dacunha2013} were taken into account in the derivation of the galaxies' intrinsic dust properties (see also \citealt{Jin2019}).


In the DL07 case, interstellar dust is described as a mixture of carbonaceous and amorphous silicate 
grains, whose size distributions are chosen to mimic different observed extinction laws. 
The result of the DL07 fit is an estimate of the dust mass and infrared luminosity;  
see \citet{dl01}, \citet{li2001}, \citet{DL07}, and \citet{berta2016} for a detailed description of the 
model and its implementation.

For both models, best-fit solutions are found in two ways: through $\chi^2$ minimization and 
through 1000 Monte Carlo (MC) realizations for each source. Uncertainties are 
computed based on $\Delta\chi^2$ or as the dispersion of all MC realizations, respectively. 
The two approaches lead to comparable results. 



\begin{table*}[!ht]
\tiny
\centering 
\caption{Infrared luminosities, dust masses, dust temperatures, and spectral emissivity indices}
\begin{tabular}{l cccccc}
\hline
\hline
Source     & $\mu_L L$(MBB) $[10^{12}$ $L_\odot]$ & $ \mu_L L$(DL07) $[10^{12}$ $L_\odot]$ & $ \mu_L M_\textrm{dust}$(MBB) & $T_\textrm{dust}$(MBB) & $\mu_L M_\textrm{dust}$(DL07) & $\beta$ \\
               &        50-1000 $\mu$m   &   8-1000 $\mu$m   &  $[10^{10}$ $M_\odot]$ & $[$K$]$ & $[10^{10}$ $M_\odot]$ \\
\hline
HerBS-34        &       24.8$\pm$0.6    &       40.6$\pm$2.4    &    1.00$\pm$0.07      &     33.0$\pm$1.2        &   1.16$\pm$0.07 & 1.87$\pm$0.08\\
HerBS-43a       &       19.8$\pm$0.8    &       33.4$\pm$4.1    &    0.53$\pm$0.05      &     34.6$\pm$1.7        &   0.69$\pm$0.04 & 1.91$\pm$0.11\\
HerBS-43b       &       10.0$\pm$1.3    &       15.0$\pm$2.0    &    0.58$\pm$0.12      &     29.1$\pm$5.1        &   0.45$\pm$0.04 & 2.21$\pm$0.39\\
HerBS-44        &       33.1$\pm$0.7    &  108.1$\pm$7.8        &    0.49$\pm$0.05      &     34.8$\pm$1.4        &   0.70$\pm$0.04 & 2.26$\pm$0.11\\
HerBS-54        &       14.9$\pm$0.6    &       23.6$\pm$1.6    &    1.33$\pm$0.16      &     26.7$\pm$1.6        &   1.25$\pm$0.12 & 2.15$\pm$0.16\\
HerBS-58        &       10.4$\pm$0.4    &       17.7$\pm$1.3    &    0.98$\pm$0.09      &     24.1$\pm$1.3        &   0.83$\pm$0.06 & 2.36$\pm$0.15\\
HerBS-70E       &       12.6$\pm$0.6    &       40.5$\pm$5.2    &    0.42$\pm$0.04      &     26.3$\pm$1.5        &   0.37$\pm$0.02 & 2.62$\pm$0.14\\
HerBS-70W       &       3.0$\pm$0.5         &   8.7$\pm$4.2    &    0.10$\pm$0.04       &     31.5$\pm$9.9        &   0.09$\pm$0.02 & 2.43$\pm$0.53\\
HerBS-79        &       10.5$\pm$0.4    &       18.1$\pm$1.2    &    0.81$\pm$0.14      &     24.0$\pm$1.5        &   0.83$\pm$0.18 & 2.52$\pm$0.19\\
HerBS-89a       &       19.3$\pm$0.8    &       29.0$\pm$1.5    &    1.40$\pm$0.09      &     27.6$\pm$1.4        &   1.35$\pm$0.06 & 2.08$\pm$0.11\\
HerBS-95E       &       7.8$\pm$0.7         &   11.4$\pm$1.9    &    0.53$\pm$0.07      &     27.2$\pm$3.6        &   0.50$\pm$0.05 & 2.22$\pm$0.33\\
HerBS-95W       &       10.9$\pm$0.8    &       16.1$\pm$1.8    &    1.00$\pm$0.09      &     24.2$\pm$2.1        &   0.77$\pm$0.04 & 2.46$\pm$0.24\\
HerBS-113       &       18.9$\pm$0.8    &       29.6$\pm$2.8    &    0.57$\pm$0.07      &     31.8$\pm$1.9        &   0.72$\pm$0.06 & 2.11$\pm$0.15\\
HerBS-154       &       25.7$\pm$1.1    &       81.3$\pm$7.3    &    0.38$\pm$0.03      &     38.2$\pm$2.3        &   0.46$\pm$0.03 & 2.10$\pm$0.14\\
HerBS-173       &       10.9$\pm$0.5    &       18.5$\pm$1.4    &    0.62$\pm$0.11      &     25.8$\pm$1.5        &   0.51$\pm$0.04 & 2.44$\pm$0.15\\
HerBS-204       &        9.9$\pm$0.4    &       16.4$\pm$1.0    &    0.85$\pm$0.15      &     21.6$\pm$0.9        &   0.45$\pm$0.04 & 2.96$\pm$0.09\\
\hline
\end{tabular}
\tablefoot{The infrared luminosities and dust masses are not corrected for amplification ($\rm \mu_L$ is the
magnification factor). Regarding the sources that are double, appropriate corrections were applied to 
estimate the flux densities of each source 
at 250, 350, and 500\,$\rm \mu m$ (see text for details). For the sources HerBS-204 and HerBS-173, we 
used the stacked continuum data (see Section~\ref{section:individual-sources}). In the case of HerBS-204, we adopt a 
redshift $z_{\rm spec}=2.37$, as the higher value ($z_{\rm spec} = 3.49$) yields a
dust temperature of $\rm 40 \, K$, which is slightly higher than the values derived for the other sources in this sample 
(see Sect.~\ref{section:individual-sources}). The MBB luminosities, dust masses, and temperatures include the effects of the CMB (see section~\ref{section:photometric-redshifts}). The quoted errors on the SED-fitting derived quantities are 1$\sigma$. } 
\label{tab:luminosities}
\end{table*}

\begin{table*}[!ht]
\tiny
\centering 
\caption{Physical properties of the galaxies}
\begin{tabular}{lcccc}
\hline
\hline
Source & $\mu_L L'_{\rm CO(1-0)}$ & $\mu_L M_{\rm H_2}$ & $\mu_L L_{\rm FIR}$ &  $\mu_L M_{\rm dust}$  \\
       & $\rm 10^{10} \, K \, km \, s^{-1} \, pc^2$ & $10^{10} \, M_\odot$ & $10^{12} \, L_\odot$ & $10^{10} \, M_\odot$ \\ 
\hline
HerBS-34   &    20.3$\pm$2.9    & 16.2$\pm$2.3      & 24.8$\pm$0.6   &  1.00$\pm$0.07     \\ 
HerBS-43a  &    38.7$\pm$5.6    & 30.9$\pm$4.4      & 19.8$\pm$0.8       &  0.53$\pm$0.05          \\
HerBS-43b  &    17.4$\pm$3.0    & 13.9$\pm$2.4      & 10.0$\pm$1.3   &  0.58$\pm$0.12     \\
HerBS-44   &    41.6$\pm$4.2    & 33.3$\pm$3.3      & 33.1$\pm$0.7   &  0.49$\pm$0.05     \\
HerBS-54   &    26.9$\pm$2.5    & 21.5$\pm$2.0      & 14.9$\pm$0.6   &  1.33$\pm$0.16     \\ 
HerBS-58   &    25.2$\pm$4.7    & 20.1$\pm$3.7      & 10.4$\pm$0.4   &  0.98$\pm$0.09     \\ 
HerBS-70E  &    10.2$\pm$2.8    &  8.2$\pm$2.3      & 12.6$\pm$0.6       &       0.42$\pm$0.04     \\
HerBS-70W  &     9.7$\pm$1.7    &  7.7$\pm$1.3      &  3.0$\pm$0.5   &  0.10$\pm$0.04     \\
HerBS-79   &    19.4$\pm$3.8    & 15.5$\pm$3.0      & 10.5$\pm$0.4   &  0.81$\pm$0.14     \\ 
HerBS-89a   &   34.4$\pm$5.2    & 27.5$\pm$4.1      & 19.3$\pm$0.8   &  1.40$\pm$0.09     \\ 
HerBS-95E  &    18.3$\pm$1.5    & 14.7$\pm$1.2      &  7.8$\pm$0.7       &  0.53$\pm$0.07          \\
HerBS-95W  &    20.9$\pm$3.4    & 16.7$\pm$2.7      & 10.9$\pm$0.8       &  1.00$\pm$0.09          \\
HerBS-113  &    17.2$\pm$3.1    & 13.7$\pm$2.5      & 18.9$\pm$0.8   &  0.57$\pm$0.07     \\ 
HerBS-154  &    58.5$\pm$5.4    & 46.8$\pm$4.3      & 25.7$\pm$1.1   &  0.38$\pm$0.03     \\ 
\hline
\end{tabular}
\tablefoot{None of the properties in this table has been corrected for gravitational magnification 
($\mu_L$ is the magnification factor). The table assumes no differential lensing between the CO and dust emission. The infrared luminosities and dust masses are those derived 
using the MBB approach (see Table \ref{tab:luminosities}). 
The gas masses are estimated using Eq.~\ref{eq:gas-mass}; see Sect.~\ref{section:CO-luminosities} 
and the footnote of Table~\ref{tab:luminosities} for details.} 
\label{tab:source-properties}
\end{table*}

\begin{figure*}[!ht]
\centering
\includegraphics[width=0.9\textwidth]{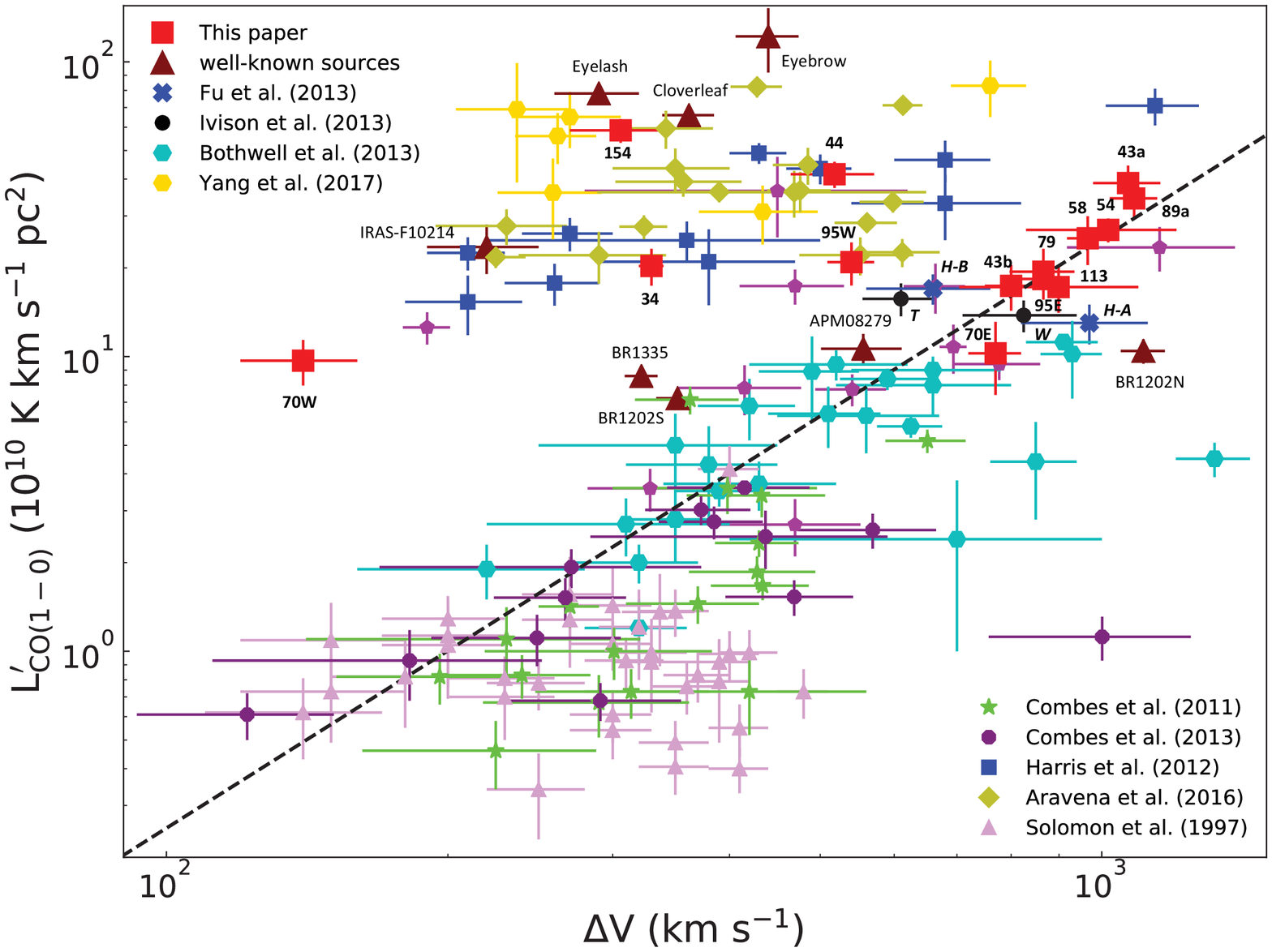}
\caption{ CO luminosity $L'_{\rm CO(1-0)}$ plotted against the linewidth 
($\Delta V$) of the CO emission line for the sources of the Pilot Program reported in this paper (large red squares),
compared to high-$z$ lensed and unlensed galaxies, as well as local ultra-luminous infrared galaxies (ULIRGs) from the literature 
with corresponding symbols as indicated in the figure. No correction for amplification
was applied to the CO luminosities. The sources of the Pilot Program are identified 
by the number of the HerBS catalogue \citep[Table~\ref{table:sources} and][]{Bakx2018}. 
Individual galaxies are also identified, namely two binary hyper-luminous galaxies (1) HATLAS~J084933 at $z=2.41$ \citep{Ivison2013},  
where the two main components, {\it W} and {\it T}, are separately labeled; 
(2) HXMM01 ({\it H-A} and {\it H-B}) at $z=2.308$ \citep{Fu2013}; 
other well-known sources, both lensed (IRASF10214, Eyelash, Cloverleaf, APM08279 and the Cosmic Eyebrow) and
unlensed (BR1202N and S, and BR1335)  from \citet[][and references therein]{Carilli-Walter2013} and \cite{Dannerbauer2019}. 
We note that the sources of \cite{Harris2012} are the ones reported in 
Table~1 of that paper, except for HATLAS~J084933 where we used the follow-up measurements of \cite{Ivison2013}. 
With the exception of the sources of this paper and the sources selected from \cite{Yang2017} and \cite{Bothwell2013}, all the other
sources used in this plot have been measured in $\rm ^{12}CO$(1-0) or, in some cases, in $\rm ^{12}CO$(2-1). 
Corrections for excitation were applied for sources for which only higher CO transitions 
are available (see Sect.~\ref{section:CO-luminosities}).
 The dashed line shows the best-fitting power-law fit derived from the data for the unlensed SMGs, 
$L'_{\rm CO(1-0)} = 10^{5.4} \times \Delta V^2$ \cite[e.g.,][]{Bothwell2013, Zavala2015}. 
}
\label{fig:Deltav-Lco}
\end{figure*}

Figure~\ref{figure:sed_fit} shows two examples of SED fits, and Table \ref{tab:luminosities} summarizes 
our findings, comparing the results of the MBB and DL07 models. For each source, the SED fits and 
the derived properties are based on the available SPIRE, SCUBA-2, and NOEMA flux densities.  

For the fields where two sources were detected within the NOEMA primary beam 
in the continuum and in at least two emission lines (namely HerBS-43, HerBS-70, and HerBS-95), the {\it Herschel} 
(even at 250\,$\rm \mu m$) and SCUBA-2 data do not provide enough information to 
separate the contributions of each component. To disentangle the flux densities, we  therefore adopted the following
methods. First, for the binary sources (HerBS-70 and HerBS-95) where the two components have the same redshift, 
we split the {\it Herschel} and SCUBA-2 flux densities 
using the average flux density ratio of the highest-frequency continuum measurements in the NOEMA 
data (see Table~\ref{tab:flux-densities}). For HerBS-70, using the average ratio of the 
flux densities at 158.6 and 149.0\,GHz shows that HerBS-70E dominates with a contribution of 82\% to 
the total flux density; for HerBS-95, the average ratio of the flux densities of the two components at 
158.6, 149.0, 143.2, and 133.5~GHz indicates that HerBS-95E contributes 41\% to the {\it Herschel} 
and SCUBA-2 flux densities. 
In the case of HerBS-43, which consists of two objects at redshift $z=3.212$ (HerBS-43a) and $z=4.054$ (HerBS-43b), we assumed a similar MBB spectrum for the two components, with a dust temperature equal to the average temperature of the whole Pilot Program sample from the MBB fits ($T_{\rm dust}$ = 30\,K; see Table~\ref{tab:luminosities}) and a dust emissivity index equal to that of the tabulated $\kappa_\nu$ ($\beta=2.08$; see \citet{li2001}, \citet{DL07}, and \citet{berta2016}).   Taking the two redshift values into account, we thus derive relative contributions of HerBS43a to the SPIRE flux densities of 85\%, 78\%, and 70\% at 250, 350, and 500\mum, respectively.

The comparison between the values of $M_\textrm{dust}$ derived from the two  
SED models (see Table~\ref{tab:luminosities}) illustrates the relative uncertainties on the dust 
mass and the models' main parameters that can be achieved with the current data. Thanks to the wide wavelength 
coverage, from $\sim$50 to $\sim$1000$\,\mu$m in the rest frame, $M_\textrm{dust}$ can be 
constrained for the majority of the sources to a 20-30\% uncertainty (3$\sigma$), 
using both the MBB and DL07 models. The $T_\textrm{dust}$ and $\beta$ are in general estimated to better than 10\%, and their averages are determined to $T_\textrm{dust}$=29$\pm$5\,K and $\beta$=2.3$\pm$0.3. It is worth noting that the Pilot Program sample is statistically too small and the dust temperatures measured too low ($T_\textrm{dust}$<60\,K) to lift the degeneracy with the spectral emissivity index $\beta$. 
The difference between the two model
estimates is a known effect \citep[e.g.,][]{berta2016}, which is mainly due to the fact that 
the MBB model is a simplification of reality and that the DL07 approach includes more dust components.
Here we  adopt the MBB results for our subsequent analysis of the sources' properties.

\subsection{Widths and profiles of the CO emission lines} 
\label{section:linewidths}
As noted previously, the distribution of the widths of the CO emission lines of 
the bright {\it Herschel} sources described in this
work is remarkable for the number of sources displaying broad lines. 
The distribution of CO linewidths is shown in 
Fig.~\ref{figure:line-widths}, where it is compared to the high-$z$ SMG samples studied by 
\cite{Bothwell2013} and \cite{Harris2012}. The mean value for the CO FWHM 
of the Pilot Program sample is $\rm 700 \pm 300 \, km s^{-1}$ with a median of $\rm 800 \, km \, s^{-1}$, 
compared to $\rm 510\pm80 \, km \, s^{-1}$ for
the unlensed SMGs from \cite{Bothwell2013}, and $\rm 525 \pm 80 \, km \, s^{-1}$ for the lensed {\it Herschel}-selected 
galaxies from \cite{Harris2012}.
 
The line profiles of the Pilot Program sources are also remarkable, as 8 out of 13 sources
display asymmetrical or double-peaked profiles with separations between the 
peaks of up to $\sim$500\,km\,s$^{-1}$, indicating
either the presence of kinematically distinct components suggestive of merger systems, or rotating disc-like components. 
Higher angular resolution observations are needed to further explore the nature of these sources.

\subsection{CO luminosities and the $L'_{\rm CO(1-0)}$ vs $\Delta V$ relationship}
\label{section:CO-luminosities}
The CO line emission traces the kinematics of the potential well in which a galaxy's molecular gas lies, and can therefore 
provide a measure of the dynamical mass of the galaxy, modulo any inclination or dispersion effects. From the integrated 
$^{12}$CO line intensity, it is possible to derive the $^{12}$CO luminosity of the source, $L'_{\rm CO(1-0)}$, which is 
related to the mass of the gas reservoir, $M_{\rm H_2}$, through  %
\begin{equation}\label{eq:gas-mass}
{M_{\rm H_2}} = \alpha L'_{\rm CO(1-0)}
,\end{equation}
where $\rm \alpha$ is a conversion factor in units of $ M_\sun \, {\rm (K \, km s^{-1} \, pc^2)^{-1}}$. 
In this paper we adopt a value of $\alpha=0.8$ suggested by measurements for SMGs and quasar hosts 
\cite[e.g.,][]{Carilli-Walter2013}. We compute the CO luminosities of the sources (in $\rm K \, km \, s^{-1} \, pc^2$) 
using the standard relation given by \cite{Solomon-VandenBout2005}, 
\begin{equation}\label{eq:CO-Luminosity}
L'_{\rm CO} = 3.25 \times 10^7 \, S_{\rm CO} \, {\Delta V \, \nu_{CO}^{-2}} \, D_{\rm L}^2 \, (1+z)^{-1}
,\end{equation}
where $S_{\rm CO}$\,$\Delta V$ is the velocity-integrated CO line flux in $\rm Jy \, km s^{-1}$, 
$\rm \nu_{CO}$ the rest frequency of the CO emission line in GHz, and $D_{\rm L}$ the luminosity
distance in Mpc in our adopted cosmology. All the CO luminosities reported in this paper 
are in $L'_{\rm CO(1-0)}$; for the sources
of the Pilot Program, we used the lowest available $J\rightarrow{}(J-1)$ transition and corrected 
for excitation adopting the median brightness temperature ratios for the SMGs 
in Table 4 of \cite{Bothwell2013}, which are compatible with the values listed in \cite{Carilli-Walter2013}, 
and applying similar corrections where needed for sources
taken from the literature (see Fig.~\ref{fig:Deltav-Lco}). Future measurements of the low-lying CO transitions 
will allow us to anchor the spectral line energy distribution for each of sources discussed in this paper
and to derive precise values for $L'_{\rm CO(1-0)}$. To homogenize the different cosmologies 
used in the various papers, we  systematically 
recalculated all $L'_{\rm CO(1-0)}$ values for the cosmology adopted in this study. 

Figure~\ref{fig:Deltav-Lco} displays the relation between the apparent CO luminosities, $L'_{\rm CO(1-0)}$,  
and the width ($\Delta V$) of the CO emission lines for the sources presented in this work and  
a compilation of high-$z$ lensed and unlensed SMGs, as well as local ULIRGs from the literature 
(see figure caption for details and references). This relationship has already been presented and 
discussed in previous studies 
    \cite[e.g.,][]{Harris2012, Bothwell2013, Carilli-Walter2013, Aravena2016, Yang2017, Dannerbauer2017, Isbell2018}. 
Figure~\ref{fig:Deltav-Lco} includes CO measurements of $\sim$160 galaxies in total; it should be noted that
none of the gravitationally lensed sources in this plot was corrected for lensing magnification. 

The most obvious feature in this figure is the clear dichotomy between the sources that are strongly lensed 
and the unlensed or weakly lensed galaxies;  the lensing magnification boosts the apparent CO luminosity 
and lifts the lensed sources above the roughly quadratic relationship between $L'_{\rm CO(1-0)}$ and 
$\Delta V$ that is observed for the unlensed sources. This trend was first pointed out by \cite{Harris2012}. 
The usefulness of this effect for measuring the lensing 
magnification has been debated in the literature, and the consensus is that deriving exact values for the magnification 
factors based on unresolved CO data is unreliable \cite[e.g.,][]{Aravena2016}. 

Notwithstanding this caution, the separation between strongly lensed and unlensed or weakly lensed sources is clearly present in the 
$L'_{\rm CO(1-0}$ versus $\Delta V$ relationship and, in principle, could be used to distinguish sources that are strongly
lensed from intrinsically hyper-luminous sources. Interestingly, the sources in the Pilot Program are located in 
the upper part of the relationship indicating that  several of the Pilot Program sources are strongly lensed, and 
that more than half of the sources are HyLIRGs located along or close to the quadratic 
relationship for the unlensed sources, akin to the hyper-luminous binary source HATLAS~J084933 
discussed in \cite{Ivison2013} and the hyper-luminous high-redshift galaxy HXMM01 
studied by \cite{Fu2013}; see identifications in Fig.~\ref{fig:Deltav-Lco}. 
The proportion of these HyLIRGs is
remarkable in this small sample selected as infrared-bright {\it Herschel} galaxies. However, more detailed observations 
are needed to verify whether these sources are lensed or not, and to derive their intrinsic properties. 
For instance, the source HerBS-89a is lensed, and a detailed discussion of its properties will be provided 
in Berta et al.\ (in preparation).

Finally, we explored the consistency of the H$_2$ masses derived independently from the $^{12}$CO and [C{\small I}]($^3$P$_1$-$^3$P$_0$) 
emission lines for the three sources where both lines were observed, namely HerBS-58, HerBS-70E, and HerBS-154. 
First, we estimated the neutral carbon masses by using Eq. (1) in \cite{Weiss2005}, assuming a [C{\small I}] 
excitation temperature equal to the derived dust temperature (see Table~\ref{tab:luminosities}). 
Adopting an atomic carbon abundance of X[CI]/X[H$_2$] of 8.4\,10$^{-5}$ \cite[]{Walter2011}, we derived 
the H$_2$ masses of HerBS-58, HerBS-70E, and HerBS-154 to be 14.2, 12.4, and 10.6\,$\times$\,10$^{10}$\,M$_\odot$, respectively. 
The comparison between the H$_2$ mass derived from the CO and the [C{\small I}] emission lines suggests that in the case of 
HerBS-70E the scaling of the high-$J$ CO line luminosity to the equivalent $J$=1-0 luminosity is most likely missing 
a significant fraction of subthermally excited gas, but that in the case of the two other sources the gas masses estimated via 
the two methods are consistent within a factor of 2. However, observations of the [C{\small I}] and 
lower-$J$ CO emission lines of a larger and statistically representative sample of SMGs are needed 
for a more accurate comparison between the two methods.


\section{Conclusion}
\label{section:conclusions}
We reported the results of a Pilot Program using NOEMA to measure reliable redshifts for a sample of 
13 bright {\it Herschel} sources. The main goal of this project was to demonstrate the ability to efficiently
derive redshifts and global properties of high-$z$ galaxies using the new correlator and broad bandwidth 
receivers on NOEMA. 
The observations described here show that the main goal of this project were successfully reached. 
Of the 13 H-ATLAS selected sources, 11 sources were detected with good signal-to-noise ratios 
 in the continuum and in at least two emission lines at 3 and 2 mm, with three sources showing an additional 
source in the field of view, allowing us to establish accurate redshifts and providing useful additional 
information on the nature and the properties of these galaxies. On average, about 100~min of total 
telescope time were needed per source in this Pilot Program to detect at least two emission lines 
in the selected $2<z<4$ bright {\it Herschel} galaxies, demonstrating the feasibility and efficiency of 
the redshift measurements using NOEMA and opening the possibility of carrying out more complete spectroscopic 
redshift surveys of larger samples of {\it Herschel}-selected galaxies.\\

The main conclusions of this paper are as follows:
\begin{itemize}
  \item Precise spectroscopic redshifts ($z_{\rm spec}$) were established for 12 galaxies 
        (including two binary systems) based on the detection of at least two emission lines, mostly 
        from $\rm ^{12}CO$ ranging from the (3-2) to the (6-5) transition. In addition, we 
        also report for three sources the detection of the emission line of the atomic carbon  
        fine-structure line $\rm [C{\small I}]\,(^3P_1$-$\rm ^3P_0)$, and in one source 
        the detection of water in the para-H$_2$O\,($2_{11}$-$2_{02}$) transition. 
        The derived spectroscopic redshifts are in the range $2.08 <z< 4.05$ 
        with a median value of $z=2.9\pm0.6$ and a tail in the distribution to $z>3$. 
  \item Combining the available continuum flux densities from {\it Herschel}-SPIRE, SCUBA-2, and the 3 and 2 mm NOEMA data, we
        assembled the SEDs of the sources and derived their infrared luminosities, dust masses, and temperatures. 
  \item The values derived for the photometric redshifts ($z_{\rm phot}$) depend on the 
        adopted photometry and the available SED coverage used in the analysis. The photometric redshifts are only indicative and are on average, for the sources studied in this paper, within 20\% of the $z_{\rm spec}$ values we  measured. 
  \item Many emission lines have broad widths between 150 and $\rm 1100 \, km \, s^{-1}$, with a mean 
        value for the CO FWHM of $\rm 700 \pm 300 \, km \, s^{-1}$ and a median of $\rm 800 \, km \, s^{-1}$.
        About 60\% of the sources display 
        double-peaked profiles indicative of merger systems and/or rotating disks.
  \item The majority of our targets are individual sources,       and are unresolved or barely resolved on 
        scales of 10\,kpc. 
        In one case, HerBS-43, there is a companion in the field, but at another redshift and hence unrelated, and in 
        the case of HerBS-89 there is a weak 2 mm continuum source, HerBS-89b, that remains undetected in line emission. Two sources (HerBS-95 and HerBS-70) are double, and in both cases the components are at the 
        same redshift with projected separation of $\sim$140\,kpc. 
  \item Based on the location of the sources studied in this paper on the $L'_{\rm CO(1-0)}$ 
        versus $\Delta V$ relationship, we conclude that several sources are gravitationally amplified, 
        and that a large fraction (including the two binaries) are hyper-luminous infrared galaxies (HyLIRGs). 
        Precise measurements of the amplification factors and the derivation of the properties of these 
        sources will require higher resolution follow-up observations in the submillimeter, and at 
        optical--near-infrared wavelengths to study the characteristics of the foreground amplifying galaxy. 
\end{itemize}

The observations presented in this study have enabled the first systematic measurement of redshifts of 
high-$z$ galaxies using NOEMA. Measurements of a larger and complete sample of galaxies selected from 
the {\it Herschel} surveys will provide a useful database for exploring in detail the properties of 
these sources and, using follow-up observations, the properties of the lensing systems in the case of 
gravitational amplification. Building  upon the success of the Pilot Program, we   started a 
comprehensive redshift survey of a sample of 125 of the brightest (S$_{500\,{\mu}\rm m}$>\,80\,mJy) 
galaxies from the {\it Herschel} surveys. The results of this ongoing survey will be reported in a forthcoming series of papers. 
This extended sample will provide, together with other already available redshift measurements, a sizeable and homogeneous 
sample of about $\sim$200 bright {\it Herschel} selected galaxies with reliable redshifts, which will allow us to 
increase the number of known lensed galaxies at the peak of cosmic evolution, to provide the largest known sample 
of HyLIRGs, and to find additional rare objects.

\begin{acknowledgements}
      This work is based on observations carried out under project numbers W17DM and S18CR with the 
      IRAM NOEMA Interferometer. IRAM is supported by INSU/CNRS (France), MPG (Germany) and IGN (Spain).
      The authors are grateful to IRAM for making this work possible. {The anonymous referee is thanked for providing useful comments. }
      A.J.B. and A.J.Y acknowledge support from the National Science Foundation grant AST-1716585. L.D. 
      acknowledges support from the ERC consolidator grant CosmicDust (PI: H. Gomez). 
      C.Y. acknowledges support from an ESO Fellowship. H.D. acknowledges financial support from the 
      Spanish Ministry of Science, Innovation and Universities (MICIU) under the 2014 Ramón y Cajal 
      program RYC-2014-15686 and AYA2017-84061-P, the later one co-financed by FEDER (European Regional Development Funds). 
      D.A.R. acknowledges support from the National Science Foundation under grant numbers AST-1614213 and AST-1910107 
      and from the Alexander von Humboldt Foundation through a Humboldt Research Fellowship for Experienced Researchers. S.J. acknowledges financial support from the Spanish Ministry of Science, Innovation and Universities (MICIU) under grant 
      AYA2017-84061-P, co-financed by FEDER (European Regional Development Funds). The National Radio Astronomy Observatory is a facility of the National Science Foundation operated under cooperative agreement by Associated Universities, Inc.
\end{acknowledgements}

\bibliographystyle{aa} 
\bibliography{references} 

\end{document}